\documentclass[conference]{IEEEtran}
\IEEEoverridecommandlockouts
\usepackage[style=ieee,url=false,doi=false,isbn=false]{biblatex}
\usepackage{amsmath,amssymb,amsfonts}
\usepackage{algorithmic}
\usepackage{graphicx}
\usepackage{textcomp}
\usepackage{xcolor}
\usepackage{braket}
\usepackage{physics}
\usepackage{subfigure}
\def\BibTeX{{\rm B\kern-.05em{\sc i\kern-.025em b}\kern-.08em
    T\kern-.1667em\lower.7ex\hbox{E}\kern-.125emX}}

\begin{document}

\title{
  Linearly simplified QAOA parameters \\ and transferability
  \thanks{
    This work was performed for Council for Science, Technology and Innovation (CSTI), Cross-ministerial Strategic Innovation Promotion Program (SIP), ``Promoting the application of advanced quantum technology platforms to social issues'' (Funding agency: QST).
  }
}

\makeatletter
\newcommand{\linebreakand}{%
\end{@IEEEauthorhalign}
\hfill\mbox{}\par
\mbox{}\hfill\begin{@IEEEauthorhalign}
}
\makeatother

\author{
  \IEEEauthorblockN{
    Ryo Sakai\IEEEauthorrefmark{1},
    Hiromichi Matsuyama\IEEEauthorrefmark{1},
    Wai-Hong Tam\IEEEauthorrefmark{1},
    Yu Yamashiro\IEEEauthorrefmark{1},
    and Keisuke Fujii\IEEEauthorrefmark{2}\IEEEauthorrefmark{3}\IEEEauthorrefmark{4}
  }
  \IEEEauthorblockA{
    \IEEEauthorrefmark{1}\textit{Jij Inc.}, 1-4-6 Nezu, Bunkyo, Tokyo 113-0031, Japan
  }
  \IEEEauthorblockA{
    \IEEEauthorrefmark{2}\textit{School of Engineering Science, Osaka University}, 1-3 Machikaneyama, Toyonaka, Osaka 560-8531, Japan
  }
  \IEEEauthorblockA{
    \IEEEauthorrefmark{3}\textit{Center for Quantum Information and Quantum Biology, Osaka University}, \\ 1-2 Machikaneyama, Toyonaka, Osaka 560-0043, Japan
  }
  \IEEEauthorblockA{
    \IEEEauthorrefmark{4}\textit{RIKEN Center for Quantum Computing (RQC)}, 2-1 Hirosawa, Wako, Saitama 351-0198, Japan
  }
  \IEEEauthorblockA{
    r.sakai@j-ij.com,
    h.matsuyama@j-ij.com,
    w.tam@j-ij.com,
    y.yamashiro@j-ij.com,
    and fujii@qc.ee.es.osaka-u.ac.jp
  }
}

\maketitle

\begin{abstract}
  Quantum Approximate Optimization Algorithm (QAOA) provides a way to solve combinatorial optimization problems using quantum computers.
  QAOA circuits consist of time evolution operators by the cost Hamiltonian and of state mixing operators,
  and embedded variational parameter for each operator is tuned so that the expectation value of the cost function is minimized.
  The optimization of the variational parameters is taken place on classical devices while the cost function is measured in the sense of quantum.
  To facilitate the classical optimization, there are several previous works on making decision strategies for optimal/initial parameters and on extracting similarities among instances.
  In our current work, we consider simplified QAOA parameters that take linear forms along with the depth in the circuit.
  Such a simplification, which would be suggested from an analogy to quantum annealing, leads to a drastic reduction of the parameter space from 2p to 4 dimensions with the any number of QAOA layers p.
  In addition, cost landscapes in the reduced parameter space have some stability on differing instances.
  This fact suggests that an optimal parameter set for a given instance can be transferred to other instances.
  In this paper we present some numerical results that are obtained for instances of the random Ising model and of the max-cut problem.
  The transferability of linearized parameters is demonstrated for randomly generated source and destination instances,
  and its dependence on features of the instances are investigated.
\end{abstract}


\section{Introduction}
\label{sec:intro}

Quantum Approximate Optimization Algorithm (QAOA)~\cite{Farhi:2014ych,Blekos:2023nil} is an algorithm to approach combinatorial optimization problems using quantum computers~\cite{Abbas:2023agz}.
This can be regarded as a quantum-classical hybrid algorithm;
one iteratively measures the cost Hamiltonian on a variationally parametrized quantum circuit and tunes the parameters on a classical computer so that the expectation value of the cost Hamiltonian is minimized.
Once the optimal parameters are obtained, the solution of the problem is expressed as the lowest energy state (or the energy itself) of the cost Hamiltonian.

QAOA has some advantages compared to other variational quantum algorithms like VQE~\cite{Peruzzo:2013bzg},
\textit{e.g.} relatively smaller number of parameters and problem dependent construction of ansatz state.
Taking the optimization is, however, not a trivial task, and there is an arbitrariness in optimizing the parameters for a given Hamiltonian and circuit.
As a typical situation, the expression power of QAOA circuit depends on the number of layers (whose detail is given later), but increasing layers often makes the parameters being trapped at local optimal and causes vanishing gradient.
Against this fact, there are several approaches to seek for the decision strategy of the parameters in both theoretical and heuristic sense~\cite{PhysRevA.97.022304,Hadfield:2021oaf,Sack2021quantumannealing,Shaydulin:2019vsv,Streif_2020,Lee:2021hxh}.

As well as the decision strategy of the parameters, there are some motivations to consider transferring a set of parameters tuned for a given instance to other instances~\cite{Galda:2021jmn,10.3389/frqst.2023.1200975,Falla:2024eye,PhysRevA.104.052419,Shaydulin:2022ehb,Sureshbabu:2023tqu}.
Indeed, some of previous works report the concentration phenomenon, where optimized parameters take similar values regardless of whatever the specific instance is~\cite{akshay2021parameter,Brandao:2018qoa,Farhi:2019xsx}.
When considering applications on real devices, transferring the variational parameters is important as an economical approach to systems with a large number of qubits.
In such systems, even taking a single sample costs much,
so that it is not tolerable to iteratively calculate the energy expectation value for the classical optimization of variational parameters,
and thus parameter transferring from a small instance to large instances will give tremendous benefits.
Also, in Reference~\cite{Mele:2022rqi}, it is pointed out that the parameter transferring is a promising approach to avoid the notorious barren plateau problem.

In another context, parameter transferring will be useful when considering relaxing constrained problems to Quadratic Unconstrained Binary Optimization problems using \textit{e.g.} the penalty method~\cite{Lucas:2013ahy}.
In the penalty method, constraints are converted to weighted penalties, where the optimal weights are not always trivial;
then one needs to take an optimization of the weights.
Therefore, when applying QAOA to such problems, the optimization of the QAOA parameters and the optimization of the weights of penalties form a nested structure, and the cost of the optimization becomes multiplicatively severe.

Motivated both to decision strategy of the parameters and to transferability, we consider simplified QAOA parameters that take linear forms along with the depth of the QAOA layers.
Indeed, an analogy to the trotterized quantum annealing suggests the linear initial QAOA parameters~\cite{Sack2021quantumannealing};
on the other hand, in the current work we fix the parameters to be linear instead of using them as initial values for the optimization.
Such simplification drastically reduces the dimension of parameter space, and, in the reduced parameter space, cost landscapes show a similarity among instances as shown in the following sections.
Moreover, although one would expect that simplifying parameters contaminates the performance of QAOA, it surprisingly turns out that there exists a set of parameters that performs well regardless of what the instance to be solved is.
This fact suggests that the optimal linear parameters have some transferability among different instances,
and indeed we check the performance of a transferred parameter set that is found for a specific instance.
It will be found in the numerical section that many instances are actually solvable by the transferred linear parameter set without any fine-tuning for each instance while the quality of parameter transferring depends on specific features of the transfer destination.

This paper is organized as follows.
In Secs.~\ref{sec:method}--\ref{sec:model} we describe QAOA and the detail of the model dealt with.
The numerical results are shown in Sec.~\ref{sec:result}, and some remarks that are mainly suggested from previous other works are given in Sec.~\ref{sec:remarks}.
Finally in Sec.~\ref{sec:summary} we summarize this paper and give future outlook.

\section{Method}
\label{sec:method}

\subsection{Quantum Approximate Optimization Algorithm}
\label{sec:qaoa}

We assume we want to minimize a given cost function $C\left( z \right)$ that takes an $n$-length bit string $z=z_{1}z_{2}\cdots z_{n}$ as an argument.
In principle, one can construct a Hamiltonian whose expectation value in the calculation basis denotes the cost function,
and, in this sense, solving the combinatorial optimization problem is to obtain the lowest energy state of the Hamiltonian.
Locating the lowest energy state in the Hilbert space is, however, not always a trivial task for generic Hamiltonians~\footnote{
  Rigorously speaking, locating the ground state for a $k$-local Hamiltonian belongs to QMA-hard~\cite{Kempe:2004sak}.
}.
Here, and in this paper, we do not in terminology distinguish the energy and the cost function.

Quantum Adiabatic Algorithm (QAA)~\cite{Farhi:2000ikn,Kadowaki:1998hua} is an algorithm to get the target state by considering an adiabatic evolution of time dependent Hamiltonian that starts from a trivial Hamiltonian, whose ground state is known, to the target Hamiltonian, whose ground state is unknown.
QAOA is derived from a trotterization of QAA and turns to QAA in the $p \to \infty$ limit, where $p$ is the number of QAOA layers.
Despite the equivalence in the limit, there are some known classes of problems where QAOA outperforms QAA~\footnote{
  For example, QAA does not improve the accuracy always by extending the adiabatic transition time~\cite{Crosson:2014gud}.
};
thus QAOA is mentioned as a promising approach to combinatorial problems.

A QAOA circuit is an iteration of layers that consist of a time evolution operator by the cost Hamiltonian $e^{-i \gamma_{l} C}$  and of a state mixing operator $e^{-i \beta_{l} H_{\mathrm{mix}}}$ for each, where $l$ denotes the depth of the layer.
$H_{\mathrm{mix}}$, which is called the mixer Hamiltonian, generically takes the form $H_{\mathrm{mix}} = \sum_{j} X_{j}$ so that every qubit is flipped independently~\footnote{
  The mixer Hamiltonian $H_{\mathrm{mix}}$ in the main body of text does not depend on the detail of the problem to be solved and is designed to explore the solution space blindly.
  Indeed an improvement of this part is proposed in several contexts.
  One of them is Reference~\cite{Hadfield:2017yqz}, where a problem dependent mixer Hamiltonian is designed so that the constraints in the problem are always satisfied;
  this algorithm is called Quantum Alternating Operator Ansatz.
  Another example is the Grover mixers~\cite{Bartschi:2020oez}.
}.
As the initial state, the superposition of all possible state $\Ket{+}^{\otimes n}$ is taken.
By applying $e^{-i \gamma_{l} C} e^{-i \beta_{l} H_{\mathrm{mix}}}$ repeatedly for each $l$ to the initial state, one obtains the final state $\Ket{\boldsymbol{\gamma}, \boldsymbol{\beta}}$ that depends on the parameters and can measure the expectation value of the cost function as $\Braket{\boldsymbol{\gamma}, \boldsymbol{\beta} | C | \boldsymbol{\gamma}, \boldsymbol{\beta}}$.
This procedure is iterated with tuning the variational parameters so that the expectation value converges to the minimum.
The observation of cost function is taken place on a quantum computer, and the optimization of parameters is taken place on a classical computer,
so that this algorithm is classified in quantum-classical hybrid algorithms.

Note that the elements of $\boldsymbol{\gamma}$ and $\boldsymbol{\beta}$ are independent of those for each other layer,
and then a $p$-layer QAOA circuit has $2p$ parameters.
The sufficient number of layers is not trivial,
and also there is an arbitrariness in how to choose initial parameters and how to take the optimization.
In principle, the more the number of layers is, the richer expression power the circuit has;
however increasing the number of layers makes the numerical complexity more demanding in both quantum and classical side.
Also, when having a deep circuit on a real device, there should be a large number of noise sources in the system.
Thus, in practical uses of QAOA, one needs to fix the minimum sufficient number of layers and the suitable decision strategy for the parameters $\boldsymbol{\gamma}$, $\boldsymbol{\beta}$;
this point is a motivation to consider the simplified decision strategy and transferability at once in our current work.

\subsection{Linearly simplification of QAOA parameters}
\label{sec:linearize}

As well as analytical studies conducted to find out the optimal parameters~\cite{PhysRevA.97.022304,Hadfield:2021oaf}, there are also heuristic approaches like layer by layer determination~\cite{Lee:2023mhp} and inter(extra)polation strategy~\cite{math11092176}.
While some previous works attempt to reduce the $2p$ dimensional ($\boldsymbol{\gamma}$, $\boldsymbol{\beta}$) parameter space by imposing some constraints or modifications~\cite{Wu:2023ddl}, the simplest assumption would be a linear form
\begin{align}
  \label{eq:linearized}
  \begin{split}
    \gamma_{l} & = \gamma_{\mathrm{slope}} \frac{l}{p} + \gamma_{\mathrm{intcp.}}, \\
    \beta_{l} & = \beta_{\mathrm{slope}} \frac{l}{p} + \beta_{\mathrm{intcp.}},
  \end{split}
\end{align}
where $l$ ($0$-based) is an index that denotes the depth of the layer.
In Reference~\cite{Sack2021quantumannealing}, the authors consider a linear choice of the initial parameters in an analogy to time scheduling for quantum annealing, and then they take an optimization where the resulting parameters are not necessarily linear;
on the other hand, in our current work, we constrain the depth dependence of the parameters to be strictly linear.
Under this constraint, the dimension of the parameter space is reduced from $2p$ to $4$ for the any number of layers $p$.

In Sec.~\ref{sec:result}, with taking the advantage of the reduced parameter space, we show cost landscapes, find the optimal slopes and intercepts, and investigate the transferability among randomly generated instances.

\section{Model}
\label{sec:model}

For a given undirected graph $G=(V,E)$, an Ising model~\cite{Lenz1920,Ising1925} where the weights $\pm 1$ are randomly assigned on each edge is defined by
\begin{align}
  \label{eq:isinghamil}
  H = \sum_{(i,j) \in E} J_{ij} s_{i} s_{j}.
\end{align}
Throughout this paper we consider only connected graphs.
$J_{ij}$ denotes the weight that is assigned on the edge $(i,j)$.
For the random Ising model we do not assume any biased distribution of the signs in $J$, so that the positive and the negative signs appear in the equal probability.
Here we do not include external fields to the Hamiltonian for simplicity.
$s_{i} = \pm 1$ is a spin variable on the node $i$;
therefore, the system that is described by this Hamiltonian takes $2^{\left| V \right|}$ states.

Standing away from the purely random case above, one can impose other types of condition to the distribution of signs (\textit{cf.} the Sherrington--Kirkpatrick model~\cite{Sherrington:1975zz}), and indeed if we limit the signs in $J$ to be positive, Eq.~\eqref{eq:isinghamil} coincides to the Hamiltonian of max-cut problem
\begin{align}
  \label{eq:maxcut}
  H_{\mathrm{maxcut}} = - \frac{1}{2} \sum_{(i,j) \in E} \left( 1 - s_{i} s_{j} \right)
\end{align}
up to the constant term and factor.
Thus we can consider the max-cut problem as a special case of Eq.~\eqref{eq:isinghamil}.

In later of this paper we consider finding the minimum energy state of the random Ising model by QAOA.
We simply assign each spin variable to each qubit, so that the number of qubits $n_{\mathrm{qubits}}$ in later sections is identical to the number of nodes $\left| V \right|$ of the given graph.
Also, for later use, we define the density of edges by the number of edges over the maximum possible number of edges in the graph: $d_{\mathrm{edges}} = \left| E \right|/{}_{\left| V \right|}\mathrm{C}_{2}$.

\section{Result}
\label{sec:result}

In this section, the measurements of energy (cost function) are taken place on an ideal simulator without any noise source.
As the ideal simulator we use Qulacs~\cite{Suzuki2021qulacsfast}.
Throughout this (and next) section, the number of QAOA layers is taken to be $p=8$ unless otherwise declared.

\subsection{Search for optimal linear parameters}

If we assume the variational parameters of a $p$-layer QAOA circuit take the form in Eq.~\eqref{eq:linearized}, the parameter space to be explored reduces to the four dimensional ($\gamma_{\mathrm{slope}}, \gamma_{\mathrm{intcp.}}, \beta_{\mathrm{slope}}, \beta_{\mathrm{intcp.}}$)-space regardless of the number of layers $p$.
Then, we consider finding the optimal slopes and intercepts by searching the four dimensional parameter space for a fixed instance of the random Ising model.
Here we adopt a Bayesian estimation by Optuna~\cite{10.1145/3292500.3330701} (rather than the grid search with a fixed resolution or any other optimizer) to find the parameter set that gives the minimum energy~\footnote{
  Note that one has to be careful to use the terms like ``optimal'' parameters and ``minimum'' energy
  since constraining the parameters to take the form~\eqref{eq:linearized} may limit the expression power of the circuit.
  Thus, we emphasize that the minimum energy obtained by the linearized parameters is not necessarily identical to the minimum energy that would be found by the original QAOA without any constraint.
}.

The important features that characterize an instance of random Ising model are the number of nodes $\left| V \right|$ ($=n_{\mathrm{qubits}}$) and the density of edges $d_{\mathrm{edges}}$~\footnote{
  Other important features would be the regularity and the parity of graph~\cite{Galda:2021jmn};
  however, in this section we do not limit the graphs to be regular and do not fix the parity.
}.
As the fixed instance that is used for the parameter search, we generate a graph with $(n_{\mathrm{qubits}}, d_{\mathrm{edges}})=(16,0.6)$.
With fixing the number of layers $p$ to $8$ and the number of shots to $2^{14}$, the obtained values are
\begin{align}
  \label{eq:optparams_nqubits16_pedge0.6_nlayers8}
  \begin{split}
    \gamma_{l} & = - 0.376 \frac{l}{p} - 0.165, \\
    \beta_{l} & = - 0.881 \frac{l}{p} + 0.913.
  \end{split}
\end{align}
One may notice that $\beta_{\mathrm{slope}}$ takes a close value to $- \beta_{\mathrm{intcp.}}$.
In the analogy to the annealing time schedule, it is quite natural to set $\boldsymbol{\beta}$ so that the effect of the mixer Hamiltonian vanishes at the end of circuit.

We have conducted the same experiment for instances with other choices of $(n_{\mathrm{qubits}}, d_{\mathrm{edges}})$.
As a result, while $n_{\mathrm{qubits}}$-dependence is not so obvious, we observe that the convergence of the Bayesian estimation gets worse for $d_{\mathrm{edges}}$s that are close to $1$ (complete graph).
For instances with $d_{\mathrm{edges}}=1$ the ground states have some degeneracy, so a possible scenario is that in such a situation there would be multiple quasi-optimal parameters.

\subsection{Cost landscapes in linearized parameter space}
\label{sec:landscape_numlayers8}

\begin{figure*}[t]
  \centering
  \begin{minipage}{0.32\hsize}
    \includegraphics[width=\hsize]{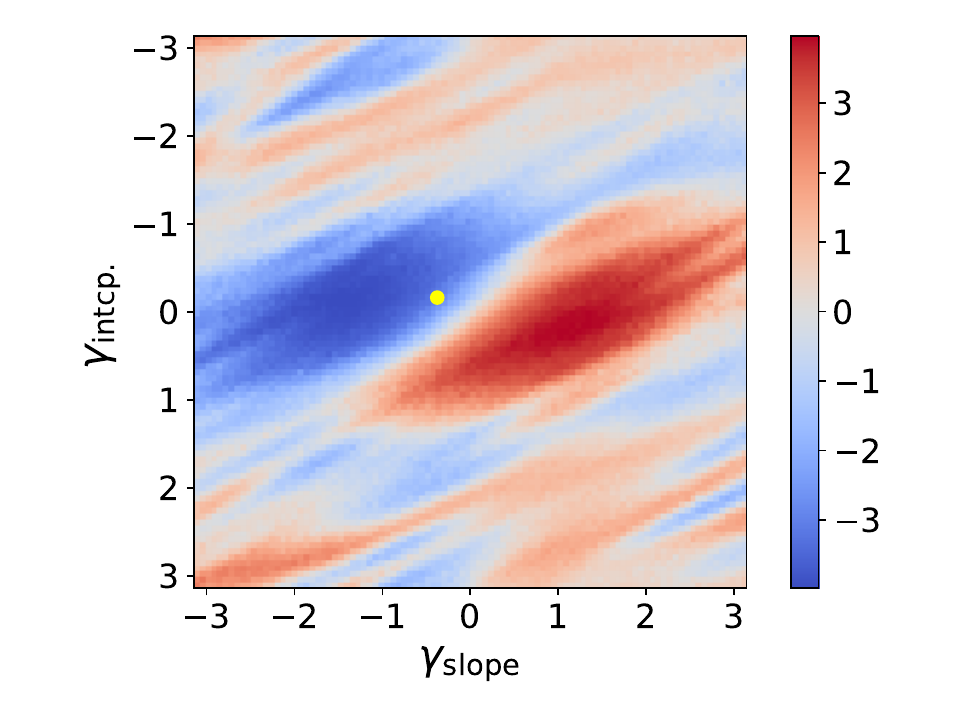}
  \end{minipage}
  \begin{minipage}{0.32\hsize}
    \includegraphics[width=\hsize]{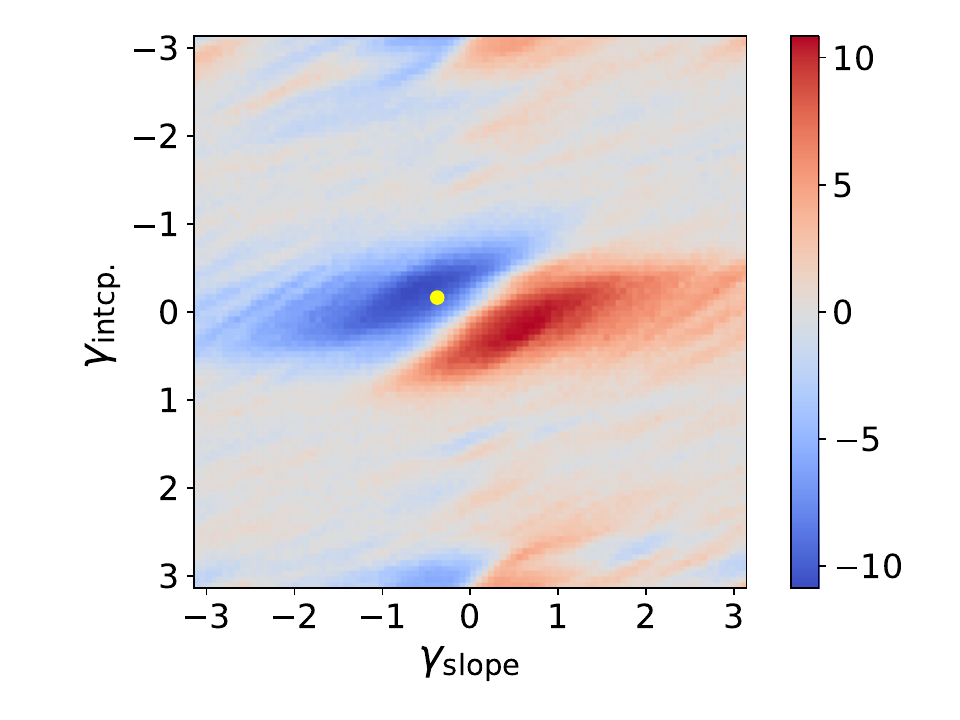}
  \end{minipage}
  \begin{minipage}{0.32\hsize}
    \includegraphics[width=\hsize]{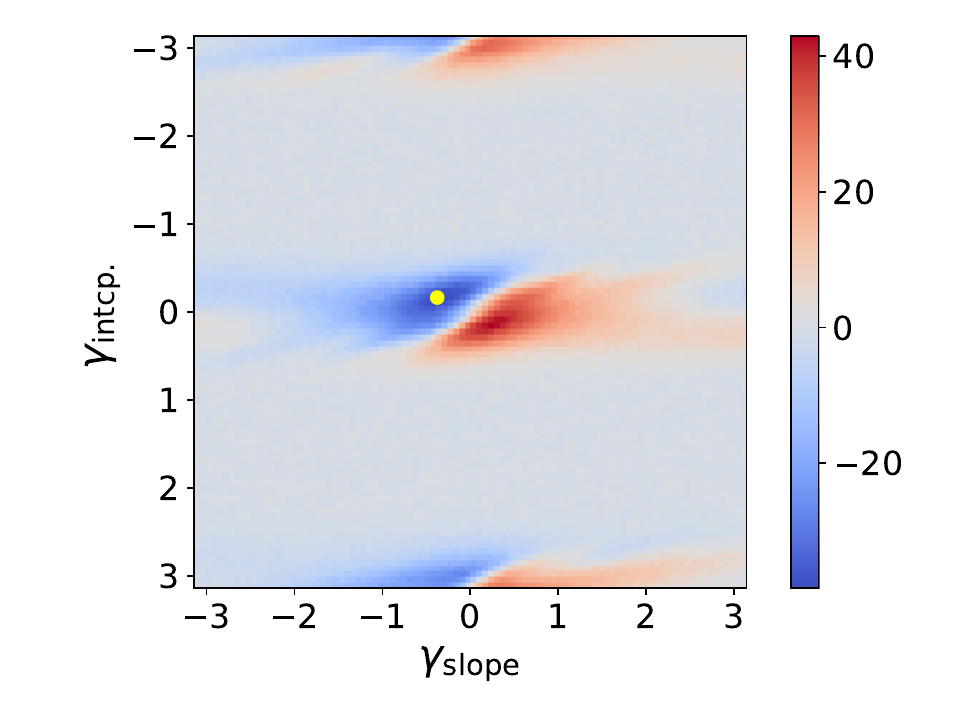}
  \end{minipage}
  \caption{
    Cost landscapes in ($\gamma_{\mathrm{slope}}, \gamma_{\mathrm{intcp.}}$)-space.
    Each panel shows the cost landscape of the Ising model on a graph that is randomly generated for $(n_{\mathrm{qubits}}, d_{\mathrm{edges}})=(5,0.42)$ (left), $(9,0.61)$ (center), $(21,0.73)$ (right).
    We point a position that corresponds to Eq.~\eqref{eq:optparams_nqubits16_pedge0.6_nlayers8} for a reference.
  }
  \label{fig:landscape_fixedbeta}
\end{figure*}

\begin{figure*}[t]
  \centering
  \begin{minipage}{0.32\hsize}
    \includegraphics[width=\hsize]{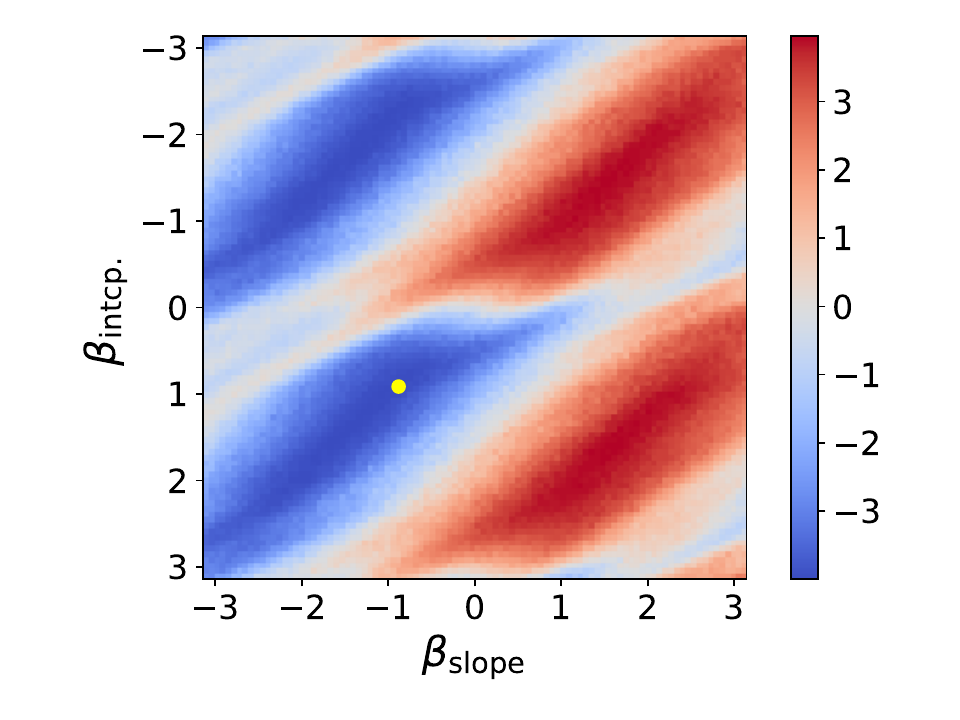}
  \end{minipage}
  \begin{minipage}{0.32\hsize}
    \includegraphics[width=\hsize]{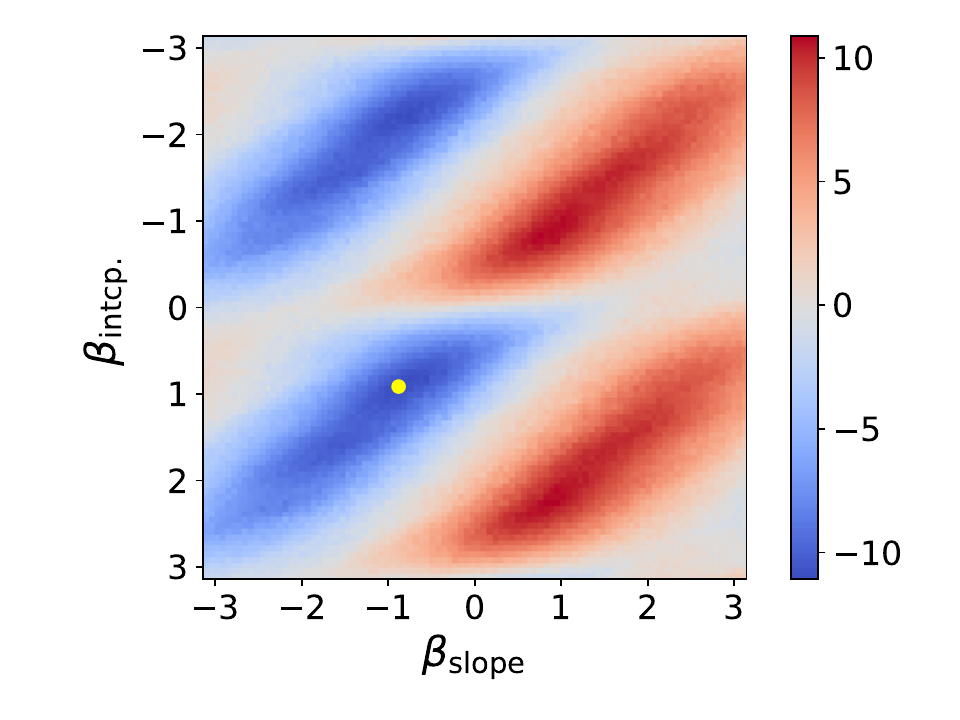}
  \end{minipage}
  \begin{minipage}{0.32\hsize}
    \includegraphics[width=\hsize]{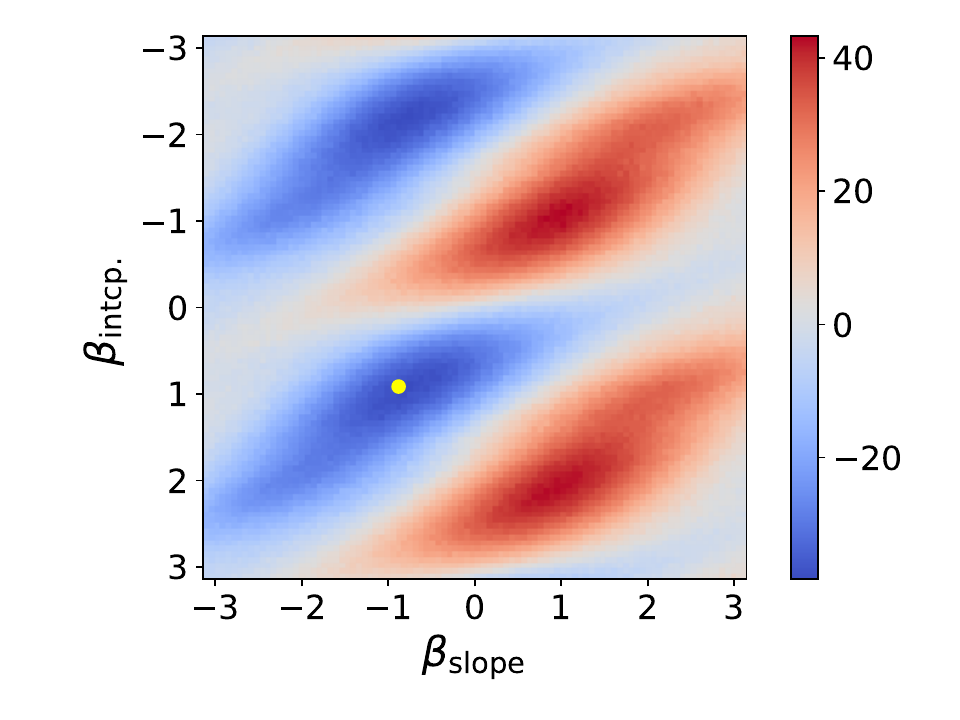}
  \end{minipage}
  \caption{
    Cost landscapes in ($\beta_{\mathrm{slope}}, \beta_{\mathrm{intcp.}}$)-space.
    Each panel shows the cost landscape of the Ising model on a graph that is randomly generated for $(n_{\mathrm{qubits}}, d_{\mathrm{edges}})=(5,0.42)$ (left), $(9,0.61)$ (center), $(21,0.73)$ (right).
    We point a position that corresponds to Eq.~\eqref{eq:optparams_nqubits16_pedge0.6_nlayers8} for a reference.
  }
  \label{fig:landscape_fixedgamma}
\end{figure*}

Here we try to discover a similarity among instances of the model by seeing cost landscapes in the linearized QAOA parameter space.
With the linearization, each of $\boldsymbol{\gamma}$ and $\boldsymbol{\beta}$ can be expressed by the two parameters (slope and intercept) regardless of the number of layers, respectively,
so that one can draw a cost landscape as a heatmap that contains all layer information for $\boldsymbol{\gamma}$ with fixing $\boldsymbol{\beta}$ and \textit{vice versa}~\footnote{
  In original QAOA without any simplification of parameters, one can draw a heatmap layer by layer with assigning $\gamma$ to an axis and $\beta$ to the other.
  Note that, on the other hand, the heatmaps in this paper contains all layer information.
}.

Figure~\ref{fig:landscape_fixedbeta} shows cost landscapes in the ($\gamma_{\mathrm{slope}}, \gamma_{\mathrm{intcp.}}$)-space.
For all panels $\boldsymbol{\beta}$ is fixed to that in Eq.~\eqref{eq:optparams_nqubits16_pedge0.6_nlayers8}~\footnote{
  Of course one can retake the Bayesian estimation for $\boldsymbol{\beta}$ for each instance instead of adopting Eq.~\eqref{eq:optparams_nqubits16_pedge0.6_nlayers8}.
  Even in that case the rough structure of landscapes is not affected much.
}.
Even though each instance is generated at random, we can observe a common structure among the instances.
Also, in Fig.~\ref{fig:landscape_fixedgamma} that shows the landscapes in the ($\beta_{\mathrm{slope}}, \beta_{\mathrm{intcp.}}$)-space in the same manner, we can find a common structure that is independent of instance as in the case of ($\gamma_{\mathrm{slope}}, \gamma_{\mathrm{intcp.}}$)-space.
The landscapes are shown for the instances with different three choices of $(n_{\mathrm{qubits}}, d_{\mathrm{edges}})$;
of course the structure is common for randomly generated instances with fixed $(n_{\mathrm{qubits}}, d_{\mathrm{edges}})$ even though they are not shown here.

Thus one can expect that the optimal set of slopes and intercepts for a instance stands close to those for other instances.
This result suggests that one can transfer the optimal set of parameters from and to each other instance.
Indeed in the next subsection we examine the transferability of the set of linear parameters in Eq.~\eqref{eq:optparams_nqubits16_pedge0.6_nlayers8} to other instances.

\subsection{Parameter transfer to other instances}

Figure~\ref{fig:xfer} shows the distribution of the expectation value of energy for randomly generated Ising models with some choices of $(n_{\mathrm{qubits}}, d_{\mathrm{edges}})$.
The energies are measured on $8$-layer QAOA circuits where the variational parameters are fixed to those in Eq.~\eqref{eq:optparams_nqubits16_pedge0.6_nlayers8}.
When one considers transferring the optimal parameters for an instance to another instance, there are two ways to do that:
(i) reusing the completely same parameters at the destination instance
and (ii) using the parameters as initial values for rerunning QAOA at the destination (imagine the transfer learning in terms of machine learning),
where ``destination'' means an instance to where a given parameter set is transferred.
On the other hand, we call an instance where an optimal set of parameters is searched ``source''.
In this paper we define the parameter transferring as just reusing the parameters without any fine-tuning for destination instances (\textit{i.e.} (i) in the above).
Thus, for the destination instance we just measure the expectation value of energy.
From the figure it is found out that there are solvable,
in the sense that the minimum energy state is most frequently observed and that the ratio of the expectation to the exact values of energy $\expval{E}/E_{\mathrm{exact}}$ is greater than $0.8$,
instances by the set of linear parameters in Eq.~\eqref{eq:optparams_nqubits16_pedge0.6_nlayers8} that is found for an instance where $(n_{\mathrm{qubits}}, d_{\mathrm{edges}})=(16,0.6)$.

\begin{figure}[htbp]
  \centering
  \includegraphics[width=\hsize]{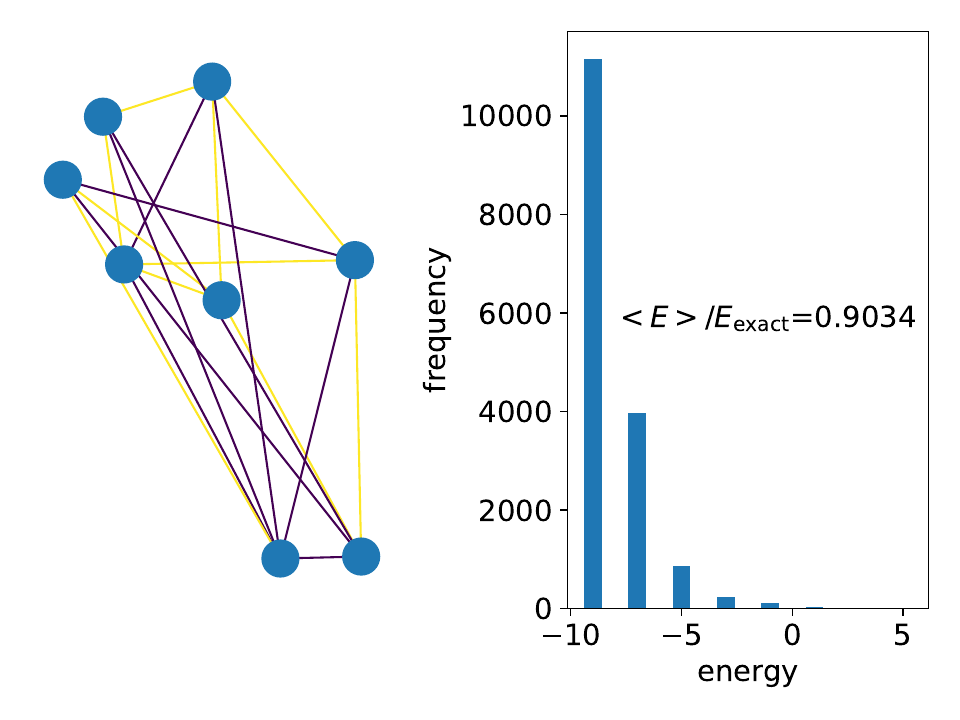}
  \includegraphics[width=\hsize]{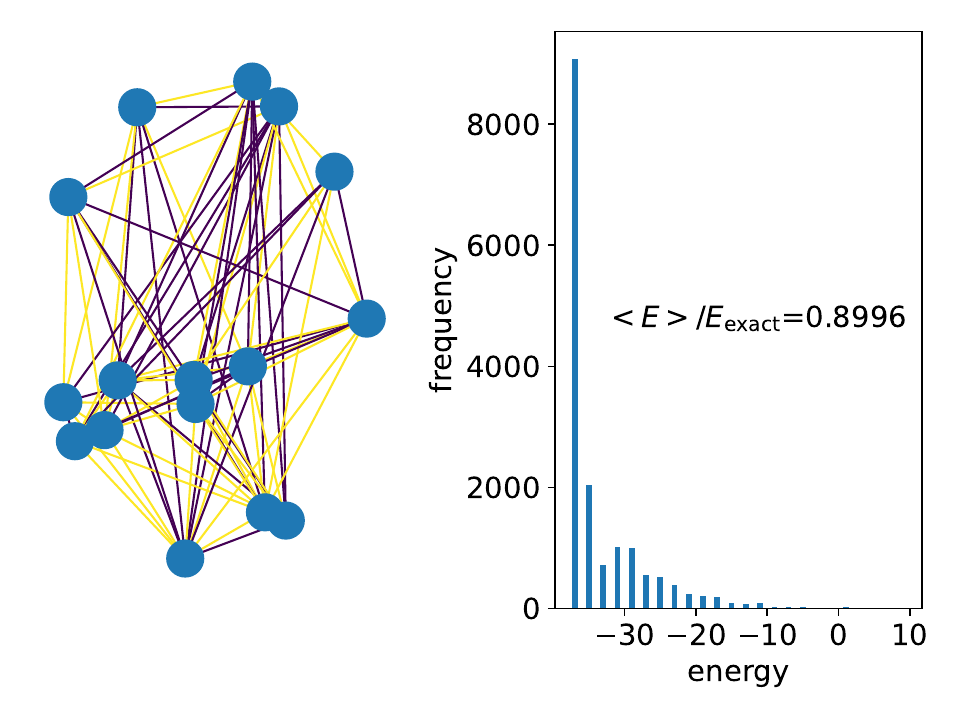}
  \includegraphics[width=\hsize]{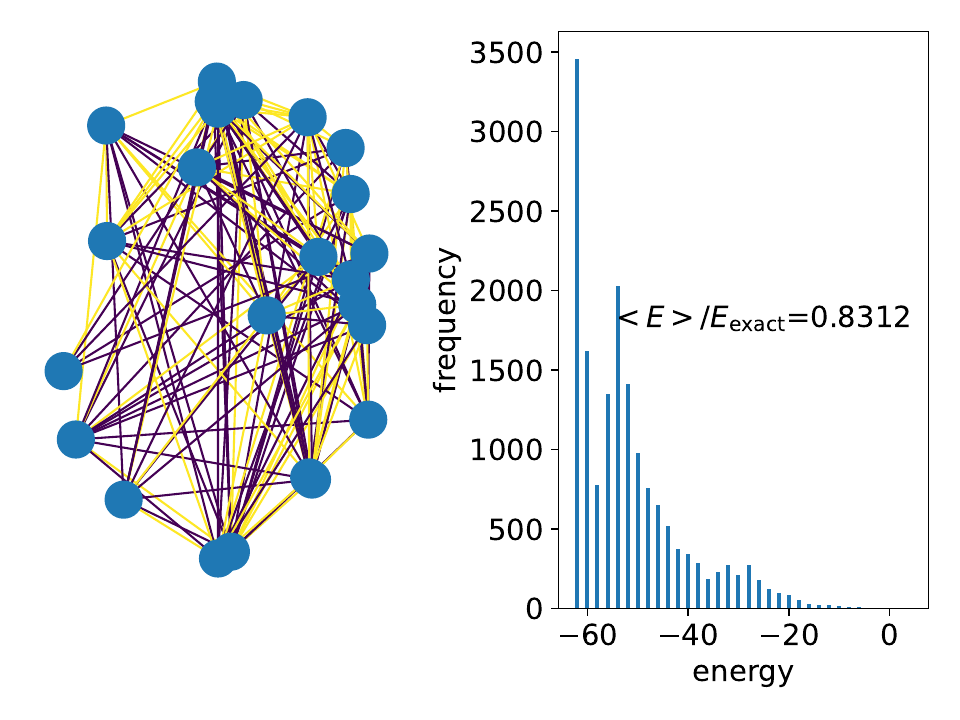}
  \caption{
    Results of transferring the parameter set in Eq.~\eqref{eq:optparams_nqubits16_pedge0.6_nlayers8} to randomly generated instances.
    $(n_{\mathrm{qubits}}, d_{\mathrm{edges}})=(8, 0.69)$ (top), $(16, 0.67)$ (middle), $(24, 0.59)$ (bottom).
    For each instance, the left panel shows the generated graph, and the right panel shows the frequency of energy value observed.
    In the graphs, the difference of edge colors corresponds to the difference of signs of weights ($+1$ or $-1$).
  }
  \label{fig:xfer}
\end{figure}

\subsection{Dependence on features of destination instance}

Figure~\ref{fig:depns_nlayers8_tunedfordedge0.6} shows $n_{\mathrm{qubits}}$- and $d_{\mathrm{edges}}$-dependence of $\expval{E}/E_{\mathrm{exact}}$.
They are scattering plots of $\expval{E}/E_{\mathrm{exact}}$ for $1024$ samples of destination instances for each panel.
From Fig.~\ref{fig:depns_nlayers8_tunedfordedge0.6}~(a), one can observe that the variance grows in the small $n_{\mathrm{qubits}}$ region.
In Fig.~\ref{fig:depns_nlayers8_tunedfordedge0.6}~(b), $\expval{E}/E_{\mathrm{exact}}$ is greater than $0.6$ for all samples in $0.1 \le d_{\mathrm{edges}} \le 1$ although the accuracy gets slightly worse around $d_{\mathrm{edges}}=0.1$.

As an extreme example for comparison, Fig.~\ref{fig:depns_nqubits16_nlayers8_tunedfordedge0.1} shows a scattering plot with varying $d_{\mathrm{edges}}$, where $\expval{E}$s are measured with a set of linear parameters $\gamma_{l} = -0.790 l/p - 0.259$, $\beta_{l} = -0.697 l/p + 0.792$ that is found for an instance where $(n_{\mathrm{qubits}}, d_{\mathrm{edges}})=(16, 0.1)$.
From this result, the worsening of transferability seems to be associated with the distance of $d_{\mathrm{edges}}$s between the source and destination instances.
In References~\cite{Shaydulin:2022ehb,Sureshbabu:2023tqu}, where transferring of generic (not linearized) QAOA parameters is studied for the weighted max-cut problem, it is claimed that one needs a shift of parameters according to the difference of energy scales between the source and destination instances.
Even in the case of random Ising model, the typical energy of a system depends on the number of edges $\left| E \right|$, so the behavior in Fig.~\ref{fig:depns_nqubits16_nlayers8_tunedfordedge0.1} seems to reflect this point.
In conclusion, the linear parameters do not always show high transferability,
and the transferability in terms of $\expval{E}/E_{\mathrm{exact}}$ can get worse depending on the features of the source and destination instances.
In such cases, cost landscapes like in Figs.~\ref{fig:landscape_fixedbeta}--\ref{fig:landscape_fixedgamma} would show some shifts of bottoms and distortions.
Indeed, one can see that zooming in and out of the $(\gamma_{\mathrm{slope}}, \gamma_{\mathrm{intcp.}})$-space changes with decreasing/increasing energy scale that would roughly correspond to the difference between the highest and lowest energies~\footnote{
  On the other hand, zooming in/out of the $(\beta_{\mathrm{slope}}, \beta_{\mathrm{intcp.}})$-space is stable.
  This would be because the elements of $\boldsymbol{\beta}$ are the coefficients of $H_{\mathrm{mix}}$ and do not depend on the problem in a sense.
}.
This point will be seen a bit more concretely in later section~\ref{sec:weighted}.

\begin{figure}[ht]
  \subfigure[$n_{\mathrm{qubits}}$-dependence. $1024$ samples are generated with fixed $d_{\mathrm{edges}}=0.6$.]{\includegraphics[width=\hsize]{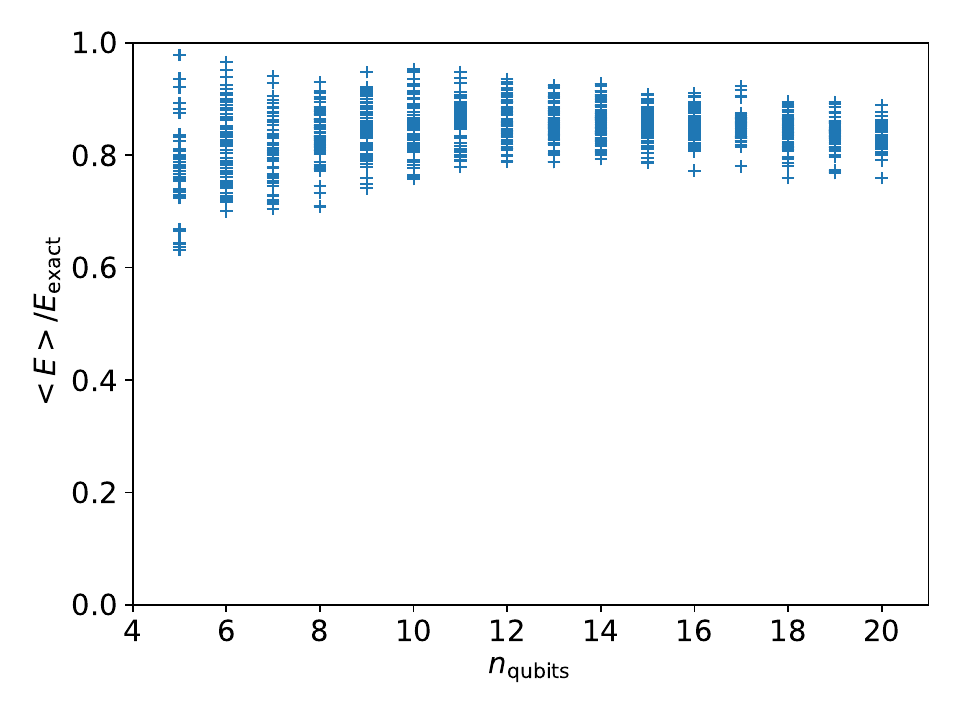}}
  \subfigure[$d_{\mathrm{edges}}$-dependence. $1024$ samples are generated with fixed $n_{\mathrm{qubits}}=16$.]{\includegraphics[width=\hsize]{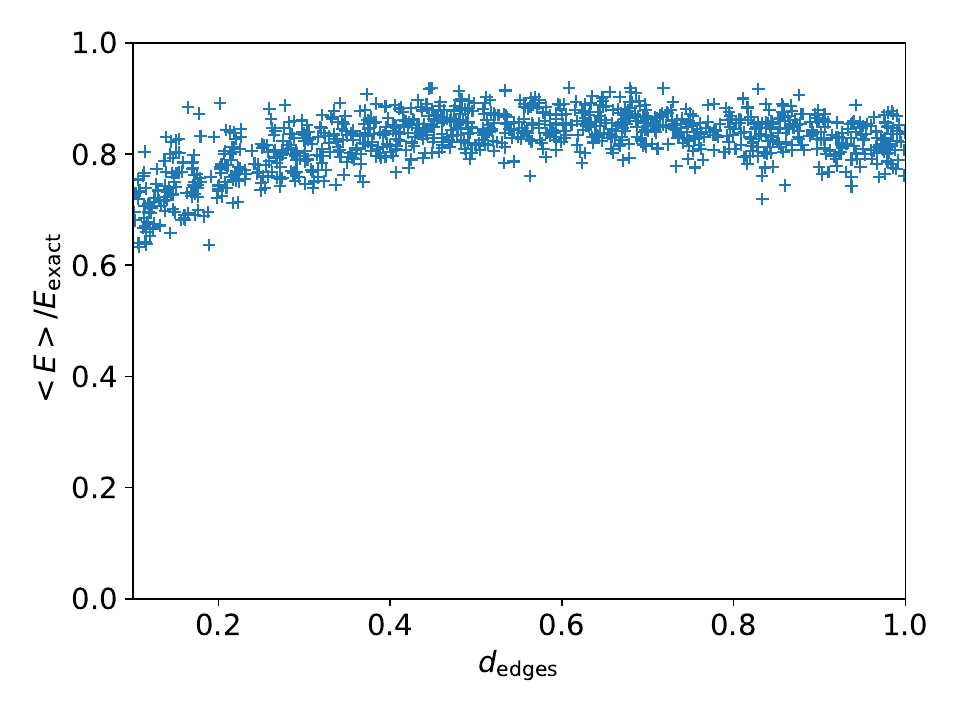}}
  \caption{
    (a) $n_{\mathrm{qubits}}$- and (b) $d_{\mathrm{edges}}$-dependence of $\expval{E}/E_{\mathrm{exact}}$.
    In both cases Eq.~\eqref{eq:optparams_nqubits16_pedge0.6_nlayers8} is used as the variational parameters in the QAOA circuits.
  }
  \label{fig:depns_nlayers8_tunedfordedge0.6}
\end{figure}

\begin{figure}[ht]
  \centering
  \includegraphics[width=\hsize]{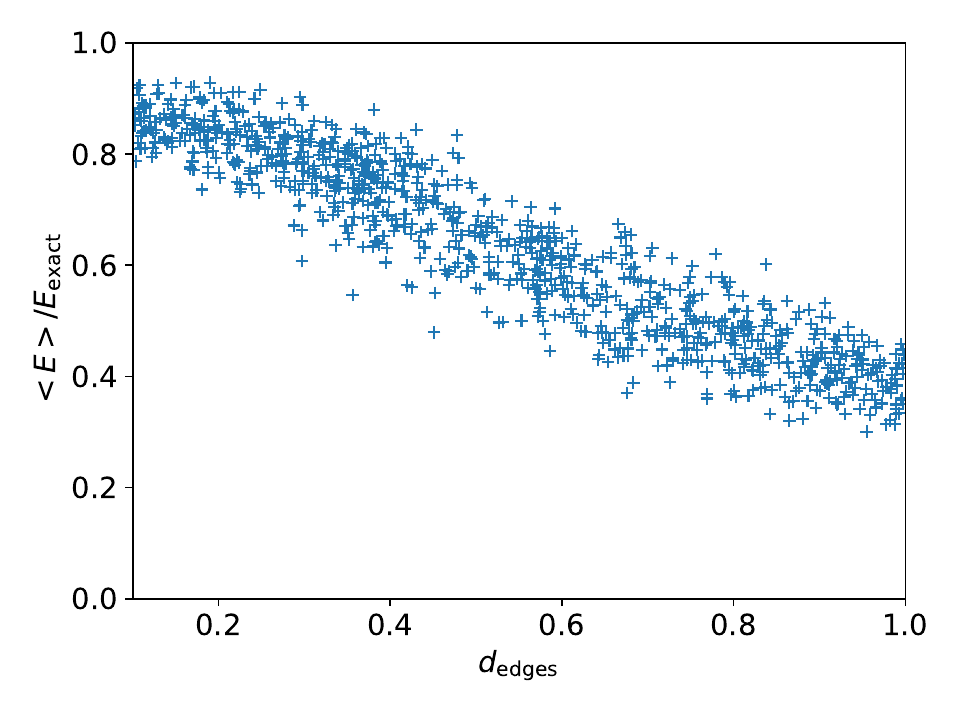}
  \caption{
    $d_{\mathrm{edges}}$-dependence of $\expval{E}/E_{\mathrm{exact}}$.
    $1024$ samples are generated with fixed $n_{\mathrm{qubits}}=16$.
    The optimal set of linear parameters found for an instance where $(n_{\mathrm{qubits}}, d_{\mathrm{edges}})=(16, 0.1)$ is used as the variational parameters in the QAOA circuits.
  }
  \label{fig:depns_nqubits16_nlayers8_tunedfordedge0.1}
\end{figure}

\subsection{Parameter transferring from random Ising model to max-cut problem}

Even though there is some instance dependent tendency, high transferability of the set of linear parameters in Eq.~\eqref{eq:optparams_nqubits16_pedge0.6_nlayers8} is demonstrated in previous subsection.
In this subsection, to check the capability of Eq.~\eqref{eq:optparams_nqubits16_pedge0.6_nlayers8} a bit more radically, we show an experiment where parameter transferring from the random Ising model to the max-cut problem is taken place.
Figure~\ref{fig:maxcut_nqubits16_rate0.68_nlayers8} shows the result of transferring Eq.~\eqref{eq:optparams_nqubits16_pedge0.6_nlayers8} to a randomly generated instance of the max-cut problem.
In this case, although the minimum energy state is not most frequently observed, $\expval{E}/E_{\mathrm{exact}}$ marks a value greater than $0.9$,
and it is suggested that there may be some transferability between problems even under different Hamiltonians.

To be a bit more concrete, we transferred Eq.~\eqref{eq:optparams_nqubits16_pedge0.6_nlayers8} to randomly generated $128$ instances of the max-cut problem where $d_{\mathrm{edges}} \in [0.4, 0.8]$ with fixed $n_{\mathrm{qubits}}=16$ and calculated the average of $\expval{E}/E_{\mathrm{exact}}$;
the result is $0.893 \pm 0.023$.

\begin{figure}[ht]
  \centering
  \includegraphics[width=\hsize]{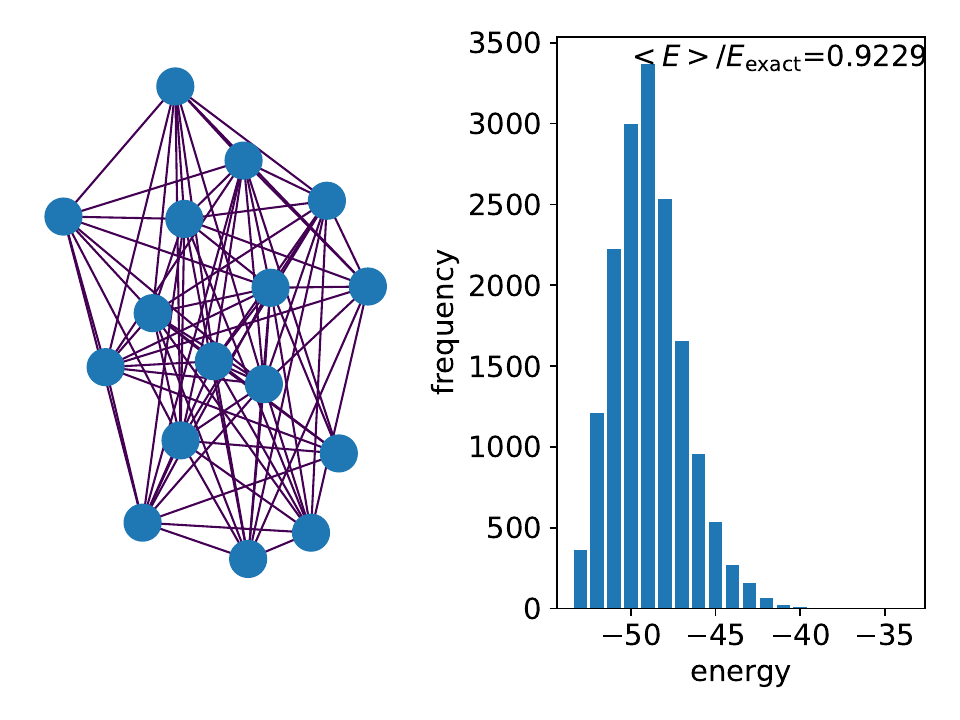}
  \caption{
    The result of transferring Eq.~\eqref{eq:optparams_nqubits16_pedge0.6_nlayers8} to an instance of the max-cut problem.
    $(n_{\mathrm{qubits}}, d_{\mathrm{edges}}) = (16, 0.68)$.
  }
  \label{fig:maxcut_nqubits16_rate0.68_nlayers8}
\end{figure}

\section{Remarks}
\label{sec:remarks}

\subsection{Dependence on energy scale}
\label{sec:weighted}

As seen in the previous section, there is a $d_{\mathrm{edges}}$-dependence of the transferability of linear parameters,
and this would be related to a dependence on the difference of energy scales between the source and destination instances.
To confirm this point, in this subsection we turn to the weighted max-cut problem $H_{\mathrm{weight}} = - (1/2) \sum_{(i,j) \in E} w J_{ij} (1 - s_{i} s_{j})$,
where each element of $J$ is uniformly sampled in $(0.1, 1)$ and where the energy scale is tunable via the weight factor $w$.

We randomly generate $1024$ instances of the weighted max-cut problem with $\left| V \right| \in [5,12]$, $d_{\mathrm{edges}} \in (0.1, 1)$, $J_{ij} \in (0.1, 1)$ $\forall \ (i,j) \in E$, and $w \in [0.1, 1, 10, 100, 1000]$,
and for each instance we estimate the optimal slopes and intercepts of $\boldsymbol{\gamma}$ and $\boldsymbol{\beta}$ by Optuna as done in the previous section.
The optimal $(\gamma_{\mathrm{slope}}, \gamma_{\mathrm{intcp.}})$ and $(\beta_{\mathrm{slope}}, \beta_{\mathrm{intcp.}})$ for each instance are shown in Fig.~\ref{fig:weightedmaxcut_optparams}.
While $(\beta_{\mathrm{slope}}, \beta_{\mathrm{intcp.}})$ does not depend on the weight factor $w$, there is a clear $w$-dependence of $(\gamma_{\mathrm{slope}}, \gamma_{\mathrm{intcp.}})$.
This tendency is somewhat natural since the elements of $\boldsymbol{\gamma}$ are the coefficients of problem Hamiltonian.
Thus a consistency of the QAOA circuit over drastic changes of the energy scale is retained by absorbing the overall factor in the Hamiltonian into $\boldsymbol{\gamma}$.
Indeed, the success ratio of QAOA $\expval{E}/E_{\mathrm{exact}}$ is not so affected and keeps being better than $0.8$ for most cases as seen in Fig.~\ref{fig:weightedmaxcut_accuracy}.

\begin{figure}[ht]
  \subfigure[Distribution of $\gamma_{\textrm{slope}}$ and $\gamma_{\textrm{intcp.}}$.]{\includegraphics[width=\hsize]{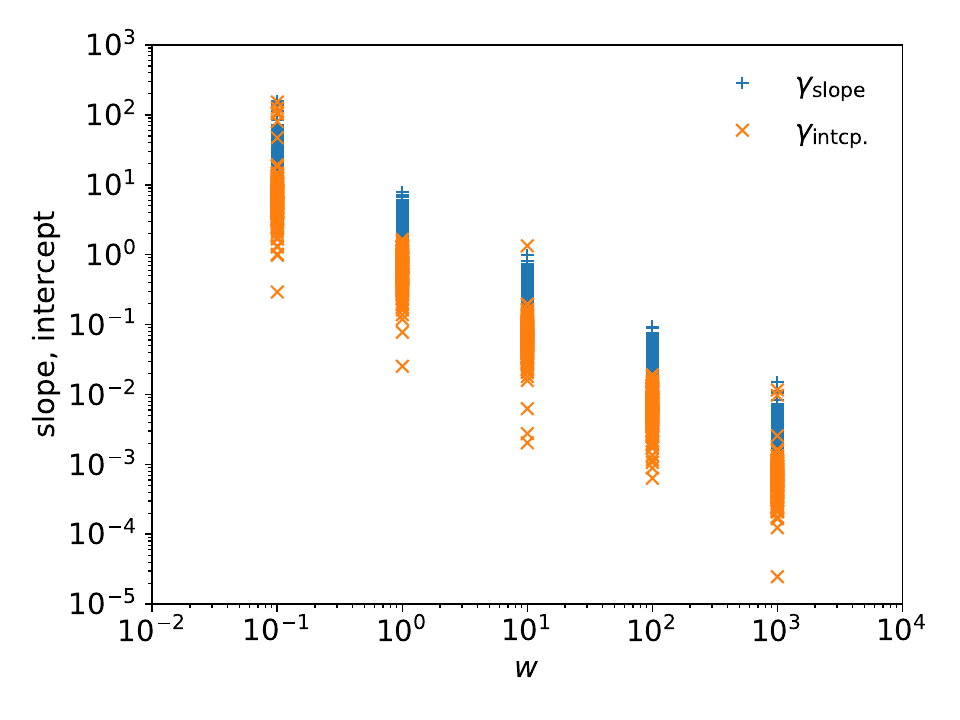}}
  \subfigure[Distribution of $\beta_{\textrm{slope}}$ and $\beta_{\textrm{intcp.}}$.]{\includegraphics[width=\hsize]{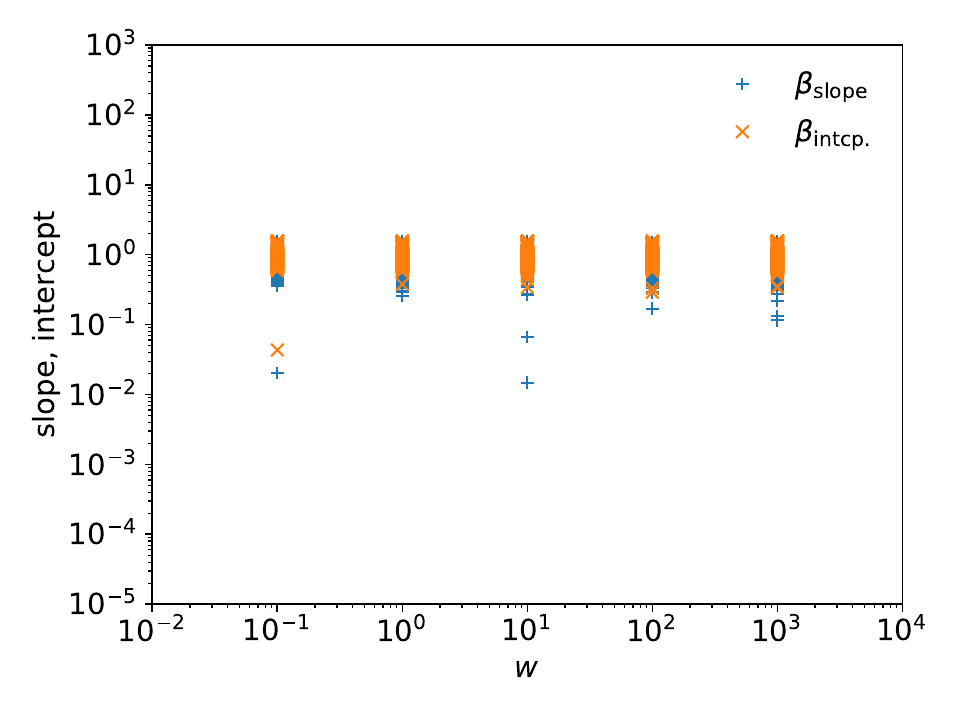}}
  \caption{
    $w$-dependence of (a) $(\gamma_{\mathrm{slope}}, \gamma_{\mathrm{intcp.}})$ and (b) $(\beta_{\mathrm{slope}}, \beta_{\mathrm{intcp.}})$.
    Note that the optimal values are searched for each instance; in other words any parameter is not transferred among them.
    $p=8$.
  }
  \label{fig:weightedmaxcut_optparams}
\end{figure}

\begin{figure}[htbp]
  \centering
  \includegraphics[width=\hsize]{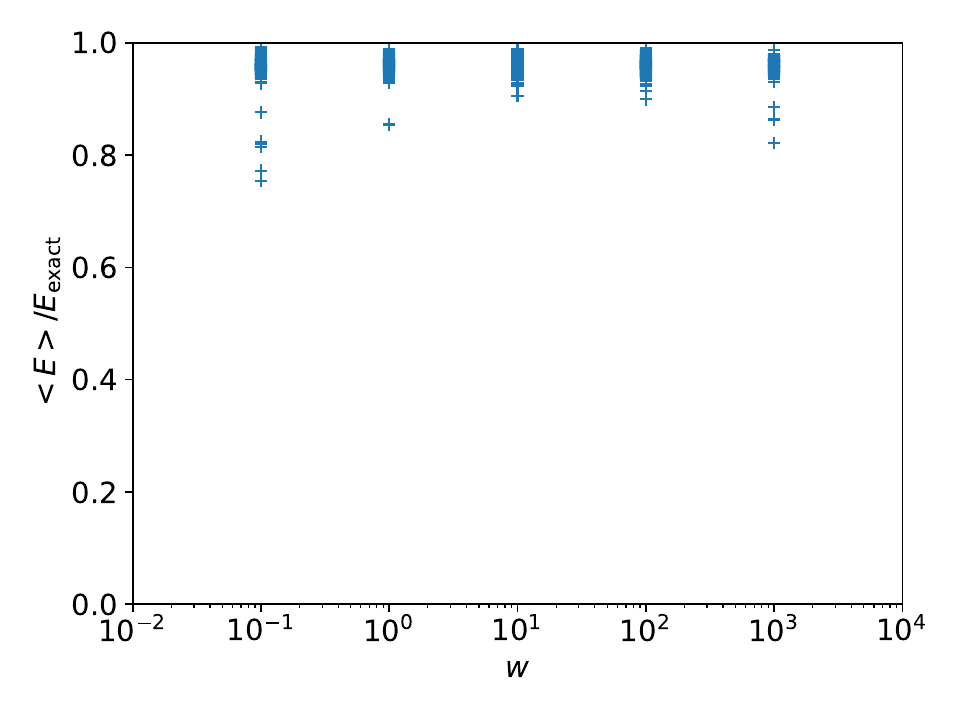}
  \caption{
    $w$-dependence of success ratio of QAOA $\expval{E}/E_{\mathrm{exact}}$.
    $p=8$.
  }
  \label{fig:weightedmaxcut_accuracy}
\end{figure}

Concerning parameter transferring, one can conclude that the shift of $\boldsymbol{\gamma}$ is required in inverse proportion to the energy scale as is also discussed in References~\cite{Shaydulin:2022ehb,Sureshbabu:2023tqu} for the generic (not linearized) QAOA.
For general models apart from the weighted max-cut problem, determining the energy scale of given system is not usually a trivial task, so that in such cases $d_{\mathrm{edges}}$ or the most significant weight among the edges would be used alternatively.

\subsection{Dependence on the number of layers $p$}

The results shown in Sec.~\ref{sec:result} are provided with the fixed number of layers $p=8$ that might be seen to be somewhat large.
In this subsection, we show cost landscapes for instances of the random Ising model with smaller choices $p=1$ and $4$.

Indeed, the parameter transferability has been intensively studied for $p=1$ QAOA in Reference~\cite{Galda:2021jmn} with putting the focus on the parity of (sub)graph.
In the paper, it is conjectured and verified for the max-cut problem that
(i) transferring between regular graphs with the same degree is successful,
(ii) transferring between regular graphs with same parity is successful,
and (iii) transferring between regular graphs with different parity is NOT successful.

Figures~\ref{fig:landscape_regular_10qubits_numlayers1}--\ref{fig:landscape_regular_10qubits_numlayers4} show the cost landscapes measured on $p=1$ and $4$ QAOA circuits, respectively.
To make a connection to the previous work by others, the instance graphs are generated with fixed degrees for each in this subsection.
In the $p=1$ case, one can find parity dependent patterns in the $(\gamma_{\mathrm{slope}}, \gamma_{\mathrm{intcp.}})$-space as pointed out in Reference~\cite{Galda:2021jmn} while the landscapes in the $(\beta_{\mathrm{slope}}, \beta_{\mathrm{intcp.}})$-space is stable.
On the other hand, in the $p=4$ case, such a parity dependent behavior disappears,
and actually the landscapes show the similar structure to those in the $p=8$ case shown in Sec.~\ref{sec:landscape_numlayers8}.

\begin{figure}[htbp]
  \subfigure[Landscape for $3$-regular graph.]{
    \begin{minipage}{0.49\hsize}
      \includegraphics[width=\hsize]{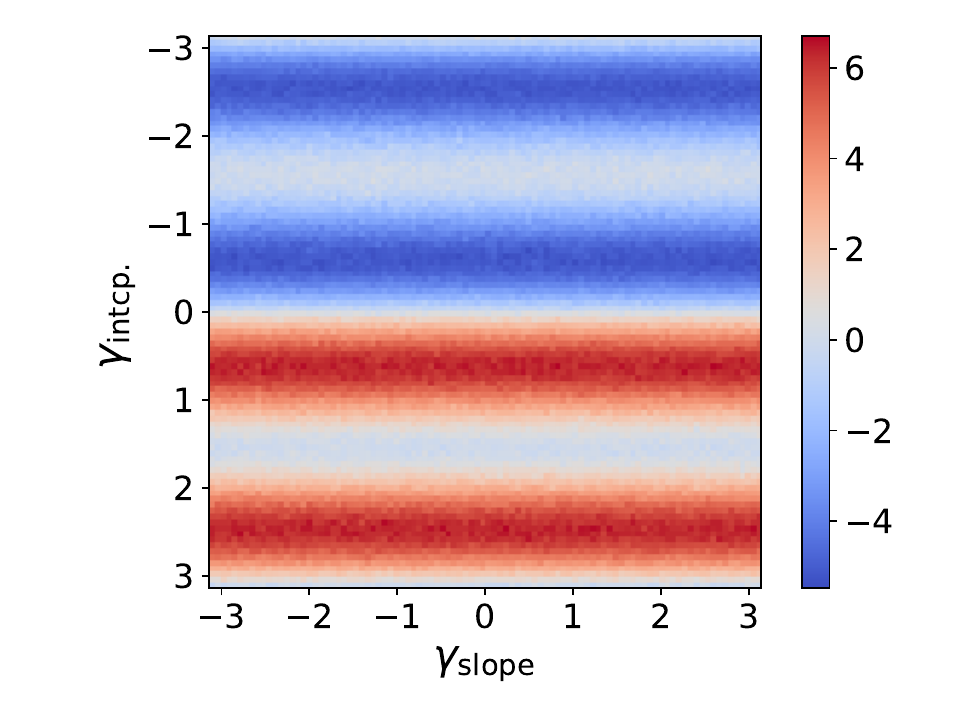}
    \end{minipage}
    \begin{minipage}{0.49\hsize}
      \includegraphics[width=\hsize]{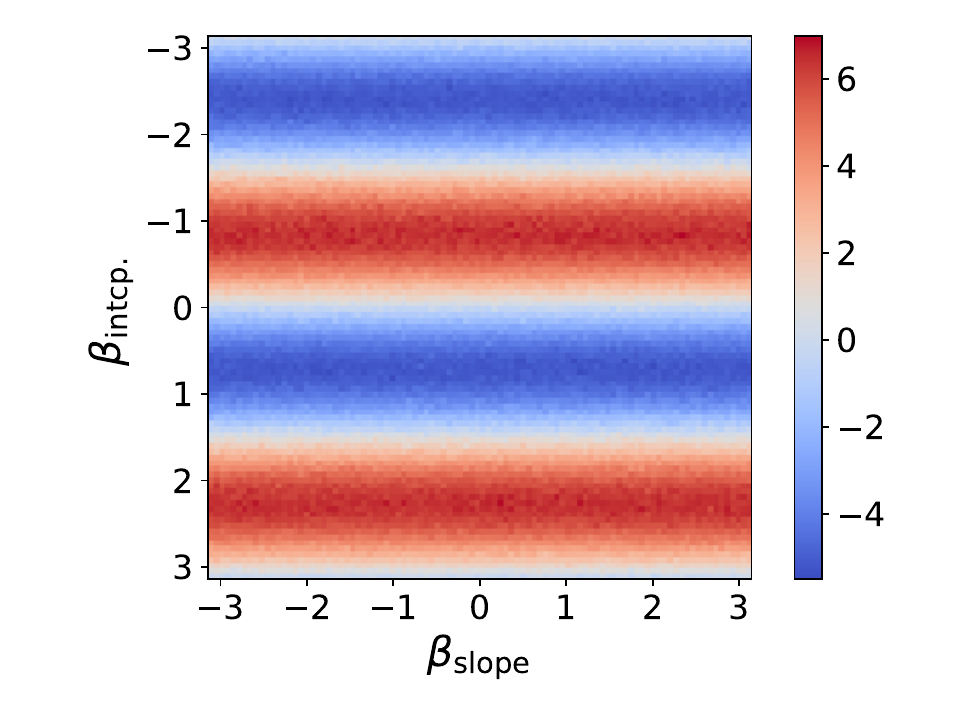}
    \end{minipage}
  }
  \subfigure[Landscape for $4$-regular graph.]{
    \begin{minipage}{0.49\hsize}
      \includegraphics[width=\hsize]{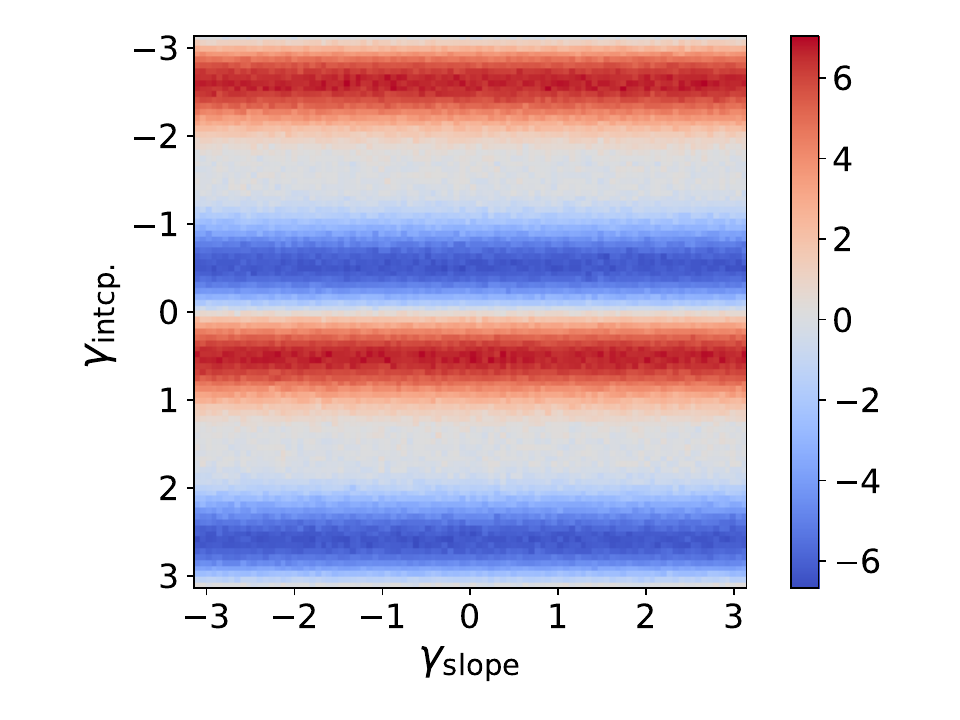}
    \end{minipage}
    \begin{minipage}{0.49\hsize}
      \includegraphics[width=\hsize]{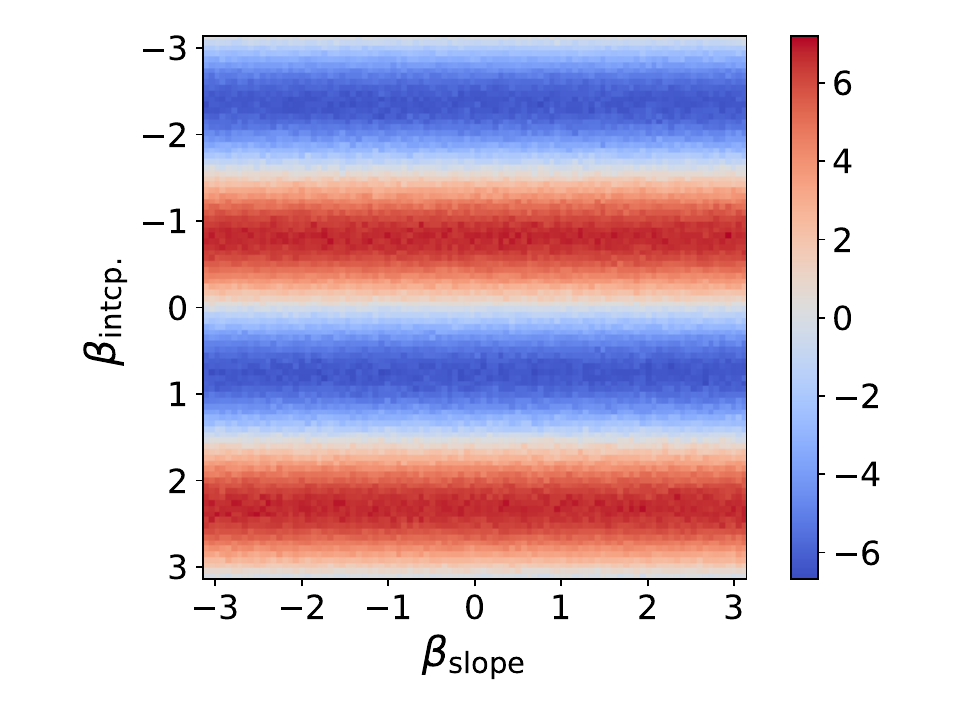}
    \end{minipage}
  }
  \subfigure[Landscape for $5$-regular graph.]{
    \begin{minipage}{0.49\hsize}
      \includegraphics[width=\hsize]{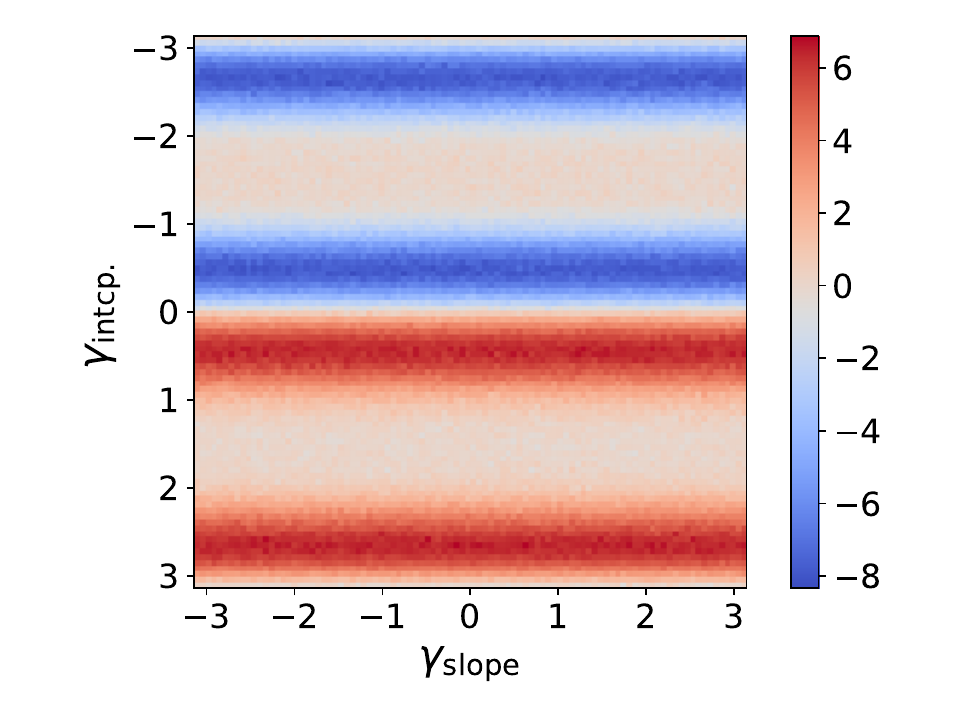}
    \end{minipage}
    \begin{minipage}{0.49\hsize}
      \includegraphics[width=\hsize]{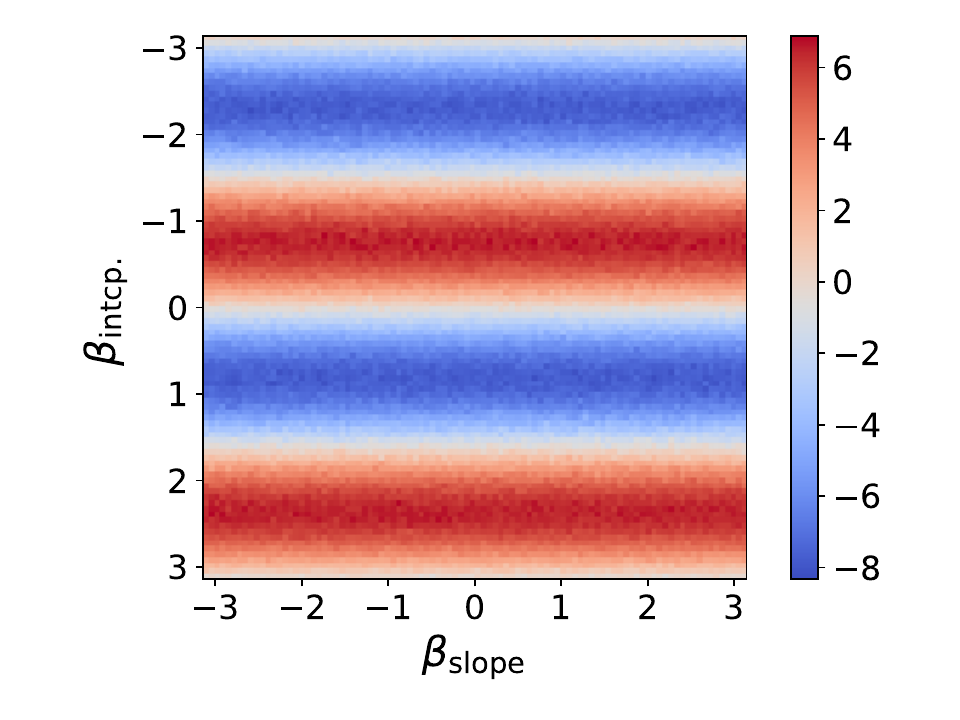}
    \end{minipage}
  }
  \subfigure[Landscape for $6$-regular graph.]{
    \begin{minipage}{0.49\hsize}
      \includegraphics[width=\hsize]{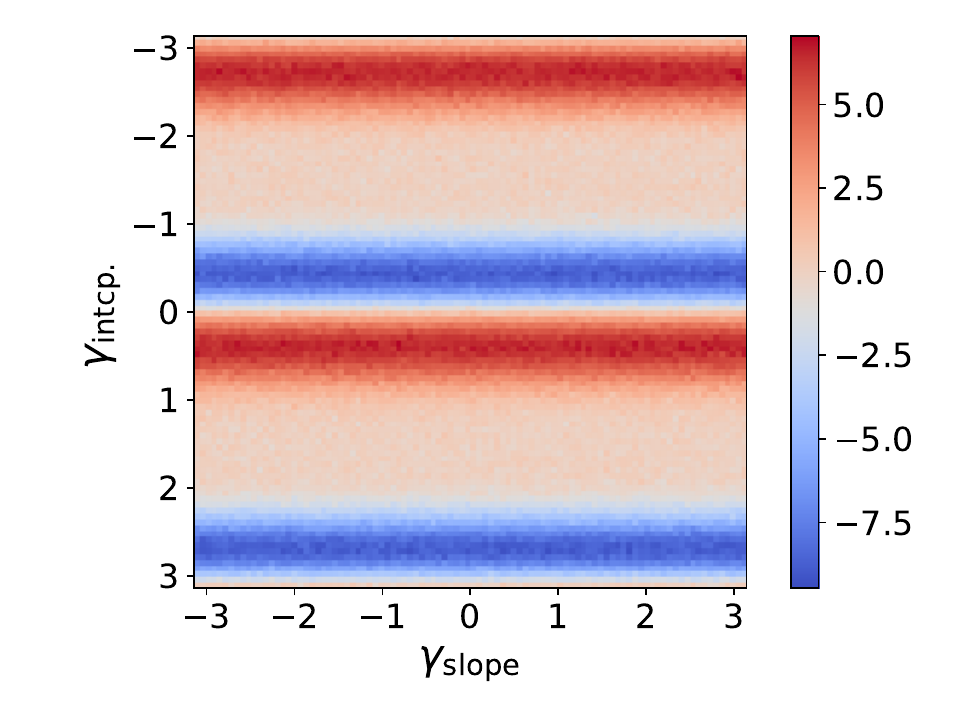}
    \end{minipage}
    \begin{minipage}{0.49\hsize}
      \includegraphics[width=\hsize]{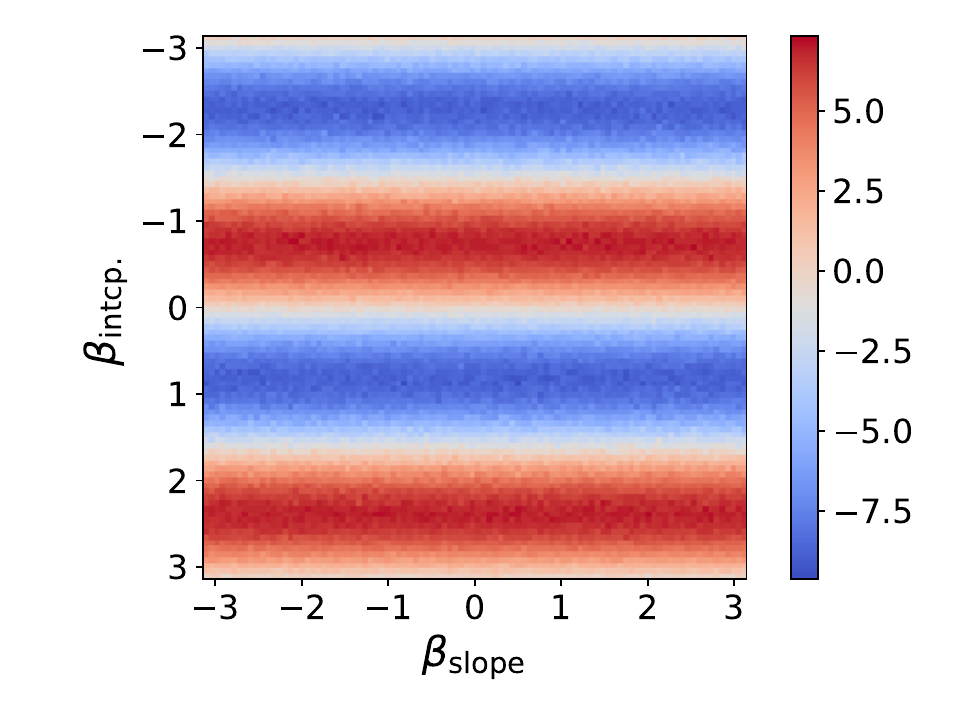}
    \end{minipage}
  }
  \caption{
    Landscapes measured on $p=1$ QAOA circuits.
    The instance graphs are chosen to be (a) $3$-, (b) $4$-, (c) $5$-, and (d) $6$-regular.
    To draw a landscape in $(\gamma_{\mathrm{slope}}, \gamma_{\mathrm{intcp.}})$-space, the fixed $\boldsymbol{\beta}$ is determined by Optuna for each instance, and \textit{vice versa}.
    Note that, in $p=1$ circuits, we cannot define the slopes of $\boldsymbol{\gamma}$ and $\boldsymbol{\beta}$, so that the horizontal axis is indeed irrelevant.
    However we show the landscapes in the same format as those with other $p$s for comparison.
  }
  \label{fig:landscape_regular_10qubits_numlayers1}
\end{figure}

\begin{figure}[htbp]
  \subfigure[Landscape for $3$-regular graph.]{
    \begin{minipage}{0.49\hsize}
      \includegraphics[width=\hsize]{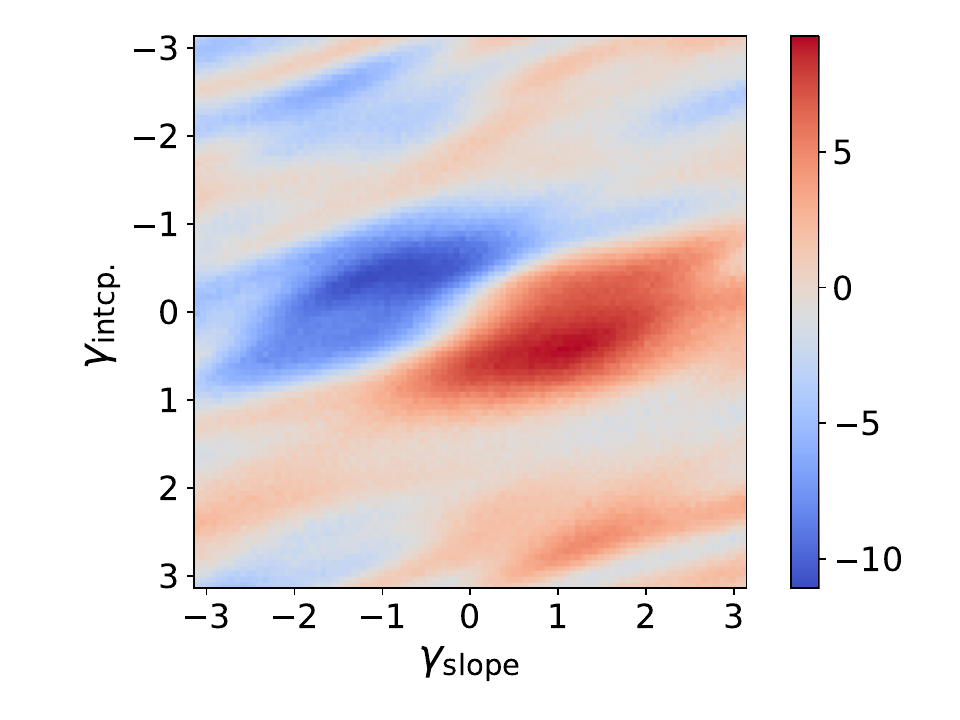}
    \end{minipage}
    \begin{minipage}{0.49\hsize}
      \includegraphics[width=\hsize]{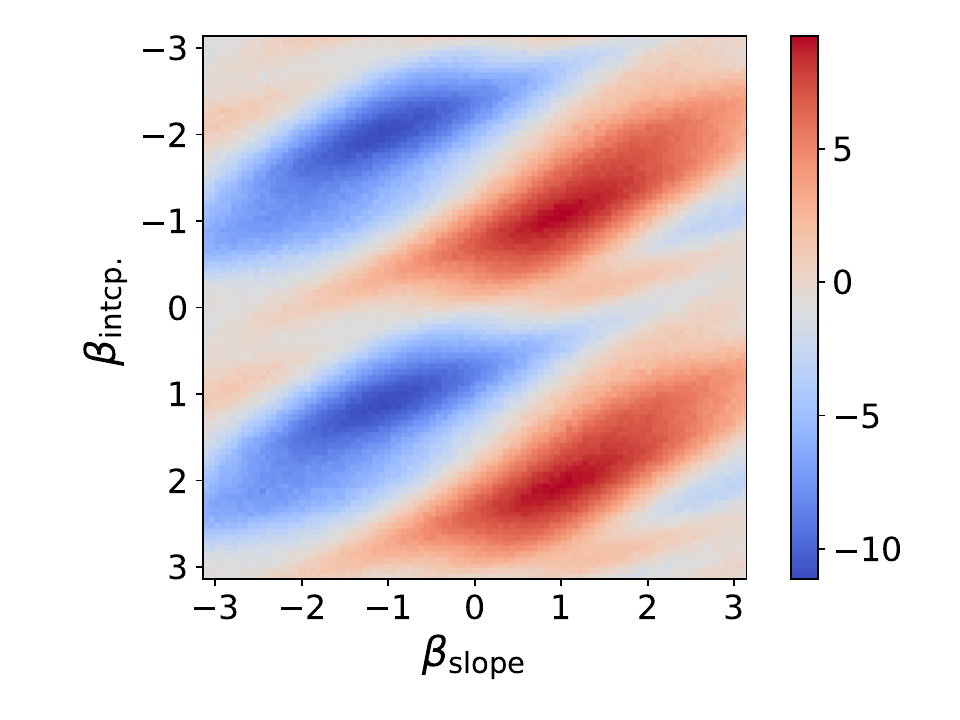}
    \end{minipage}
  }
  \subfigure[Landscape for $4$-regular graph.]{
    \begin{minipage}{0.49\hsize}
      \includegraphics[width=\hsize]{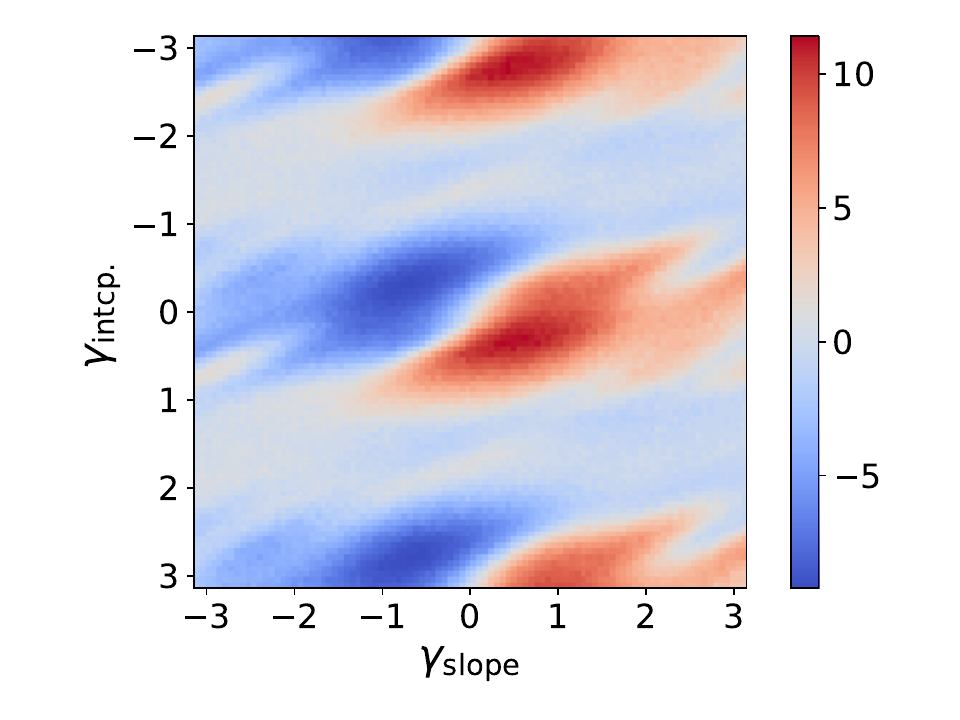}
    \end{minipage}
    \begin{minipage}{0.49\hsize}
      \includegraphics[width=\hsize]{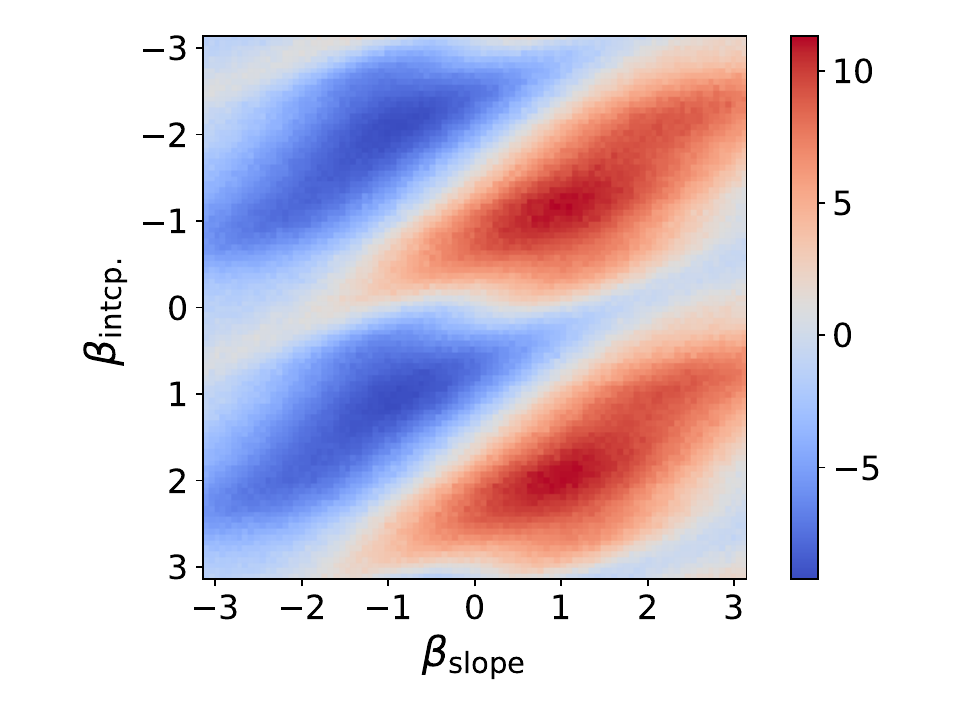}
    \end{minipage}
  }
  \subfigure[Landscape for $5$-regular graph.]{
    \begin{minipage}{0.49\hsize}
      \includegraphics[width=\hsize]{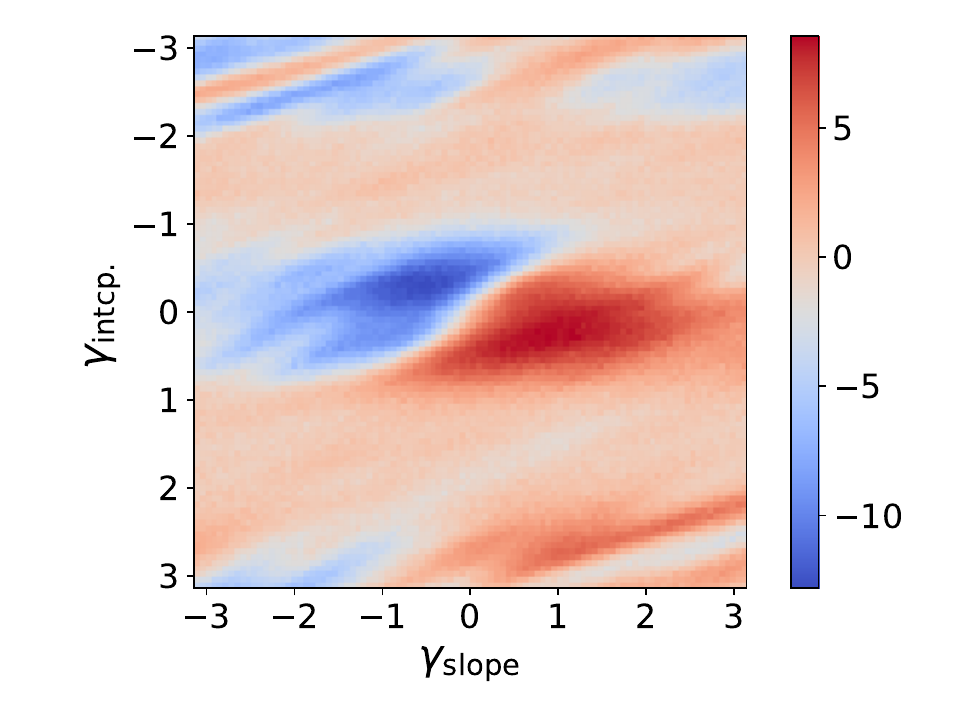}
    \end{minipage}
    \begin{minipage}{0.49\hsize}
      \includegraphics[width=\hsize]{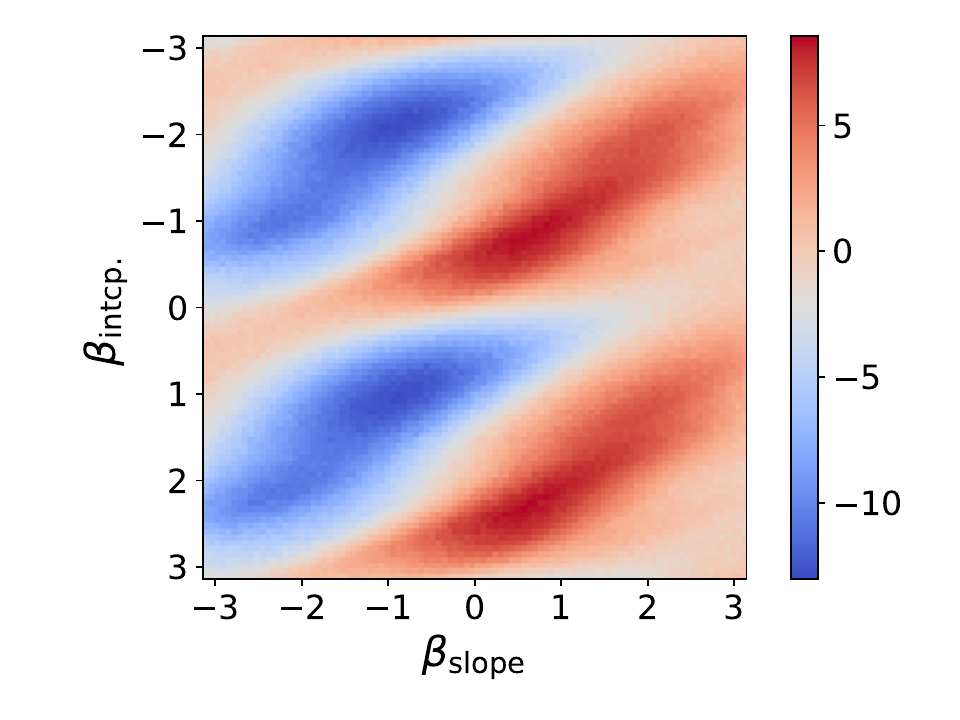}
    \end{minipage}
  }
  \subfigure[Landscape for $6$-regular graph.]{
    \begin{minipage}{0.49\hsize}
      \includegraphics[width=\hsize]{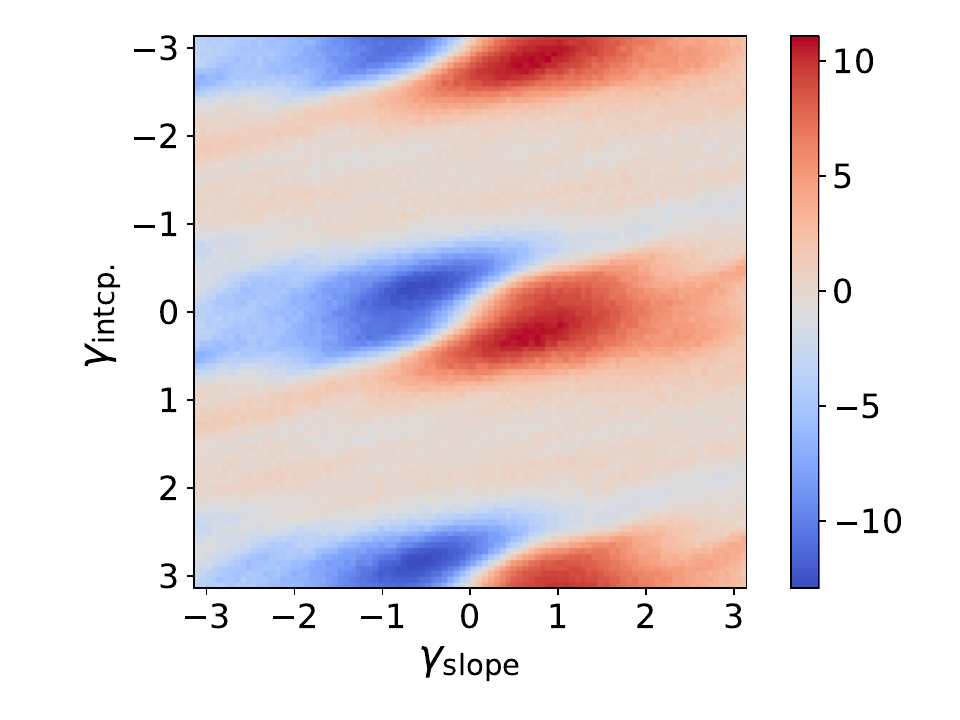}
    \end{minipage}
    \begin{minipage}{0.49\hsize}
      \includegraphics[width=\hsize]{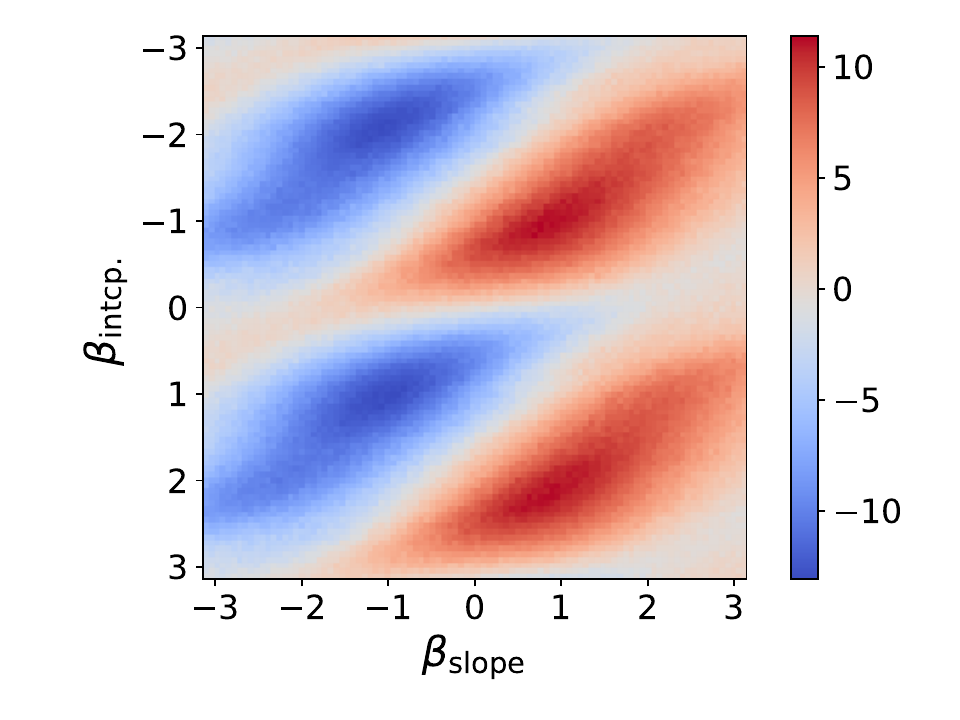}
    \end{minipage}
  }
  \caption{
    Landscapes measured on $p=4$ QAOA circuits.
    The instance graphs are chosen to be (a) $3$-, (b) $4$-, (c) $5$-, and (d) $6$-regular.
    To draw a landscape in $(\gamma_{\mathrm{slope}}, \gamma_{\mathrm{intcp.}})$-space, the fixed $\boldsymbol{\beta}$ is determined by Optuna for each instance, and \textit{vice versa}.
  }
  \label{fig:landscape_regular_10qubits_numlayers4}
\end{figure}

\subsection{Instance dependence of final states}

When transferring (fixing) parameters, a possible concern is that similar superposition states are produced regardless of the detail of instance.
To visualize the tendency of resulting superposition states of QAOA, we check the fidelity $\left| \Braket{\text{src.}|\text{dest.}} \right|$, where $\Ket{\text{src.}}$ and $\Ket{\text{dest.}}$ are the QAOA states for the source and destination instances, respectively.
$1024$ samples are generated as the destination instances, where $n_{\mathrm{qubits}}$ is fixed to be $8$~\footnote{
  Note that this means we use another parameter set instead of Eq.~\eqref{eq:optparams_nqubits16_pedge0.6_nlayers8}.
  Indeed, we have observed clear $n_{\mathrm{qubits}}$-dependence of the fidelity;
  if we take larger $n_{\mathrm{qubits}}=16$, the fidelity decreases in an order of magnitude.
  Thus we take $n_{\mathrm{qubits}}=8$ in this subsection to roughly fit the range of fidelity to $(0,1)$.
}.
On the other hand, $d_{\mathrm{edges}}$ is taken from $(0.1, 1)$ at uniformly random for the destination instances while $d_{\mathrm{edges}} = 0.6$ for the source instance.

First, for a reference, Fig.~\ref{fig:dedges_fidelity_nqubits8} shows the $d_{\mathrm{edges}}$-dependence of the fidelity.
There is not a clear $d_{\mathrm{edges}}$-dependence of the fidelity.
Note that there is a point with $\left| \Braket{\text{src.}|\text{dest.}} \right| = 1$;
this corresponds to transferring from the source to the source instance, which is shown as a reference level.

\begin{figure}[htbp]
  \centering
  \includegraphics[width=\hsize]{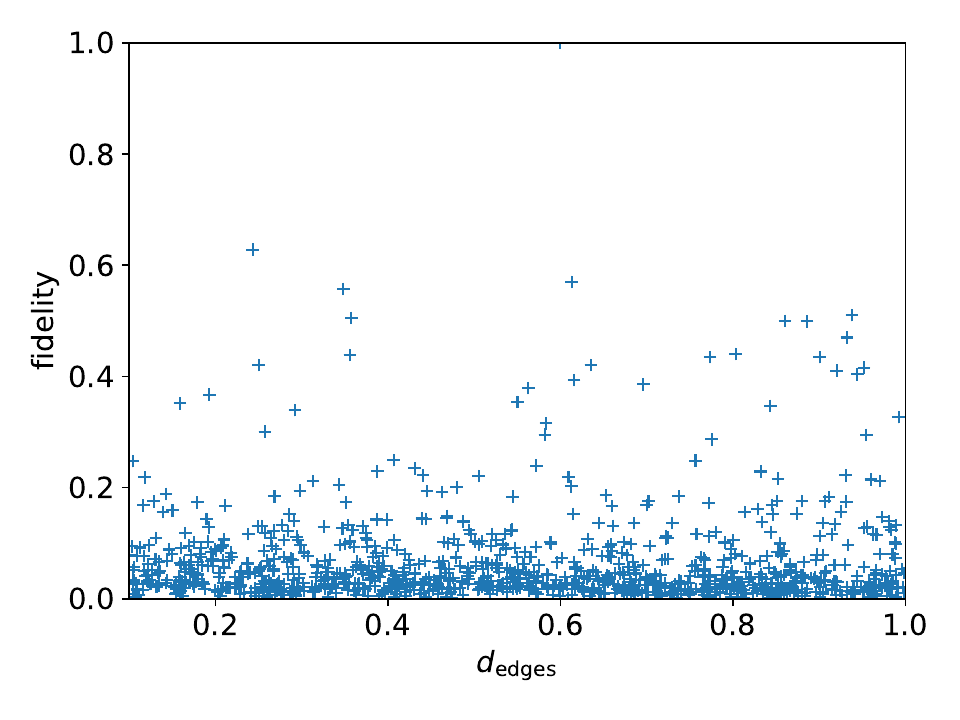}
  \caption{
    Fidelity as a function of $d_{\mathrm{edges}}$.
    $p=8$.
  }
  \label{fig:dedges_fidelity_nqubits8}
\end{figure}

Figure~\ref{fig:fidelity_ratio_nqubits8} shows the fidelity dependence of the transferability in terms of $\expval{E}/E_{\mathrm{exact}}$.
As seen from the figure, high transferability is not necessarily associated with high fidelity between the QAOA states for the source and destination instances.
Indeed, one can find the instances that mark $\expval{E}/E_{\mathrm{exact}}$ greater than that for the source instance (see the point with $\left| \Braket{\text{src.}|\text{dest.}} \right| = 1$).
To answer the concern given in the beginning of this subsection, the features of problem Hamiltonian is embedded to the QAOA circuit, and thus the amplitude of the final state depends on instance even with parameters fixed to a certain choice.

\begin{figure}[htbp]
  \centering
  \includegraphics[width=\hsize]{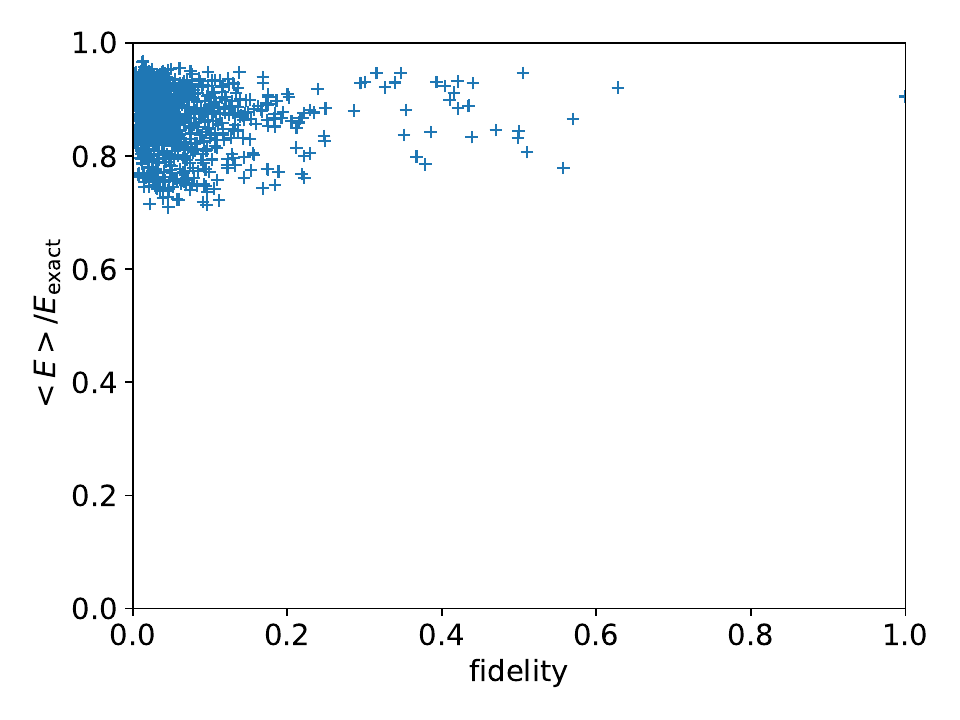}
  \caption{
    Fidelity dependence of transferability.
    $p=8$.
  }
  \label{fig:fidelity_ratio_nqubits8}
\end{figure}

\section{Summary}
\label{sec:summary}

In this paper, the behavior of linearized QAOA parameters and their transferability are investigated.
Surprisingly, cost landscapes in the (slope, intercept)-spaces take a similar shape regardless of the detail of instance.
From this fact, it is suggested that the optimal set of linear parameters found for an instance can be transferred to other instances,
and indeed we demonstrated that many instances are solvable, in terms of the distribution of observed energy states and the ratio of the expectation to the exact values of energy, without instance dependent fine-tuning.

The set of linear parameters in Eq.~\eqref{eq:optparams_nqubits16_pedge0.6_nlayers8} that is used for the experiments is found by a Bayesian estimation for an instance where $(n_{\mathrm{qubits}}, d_{\mathrm{edges}})=(16, 0.6)$.
This parameter set performs high transferability so that $\expval{E}/E_{\mathrm{exact}}$ is over $0.6$ for every destination instance, while the transferability gets worse when the difference of $d_{\mathrm{edges}}$s between the source and destination instances is large.
For example, when we transfer an optimal set of linear parameters that is found for an instance where $(n_{\mathrm{qubits}}, d_{\mathrm{edges}})=(16, 0.1)$ to an instance on the complete graph \textit{i.e.} $d_{\mathrm{edges}}=1$, $\expval{E}/E_{\mathrm{exact}}$ gets as bad as $0.3$.

While most of the experiments in this paper are taken place for the random Ising model, we have also tried transferring to the max-cut problem that can be regarded as a special case of the former model.
As a result, there are instances of the max-cut problem where the set of parameters in Eq.~\eqref{eq:optparams_nqubits16_pedge0.6_nlayers8} achieves $\expval{E}/E_{\mathrm{exact}} > 0.9$ even though Eq.~\eqref{eq:optparams_nqubits16_pedge0.6_nlayers8} is found for an instance of the random Ising model.

In this paper we exclusively use a noiseless simulator, so that the verification on real devices is remaining as a future work.
Simplifying parameters and not taking instance dependent fine-tuning, we, in a sense, ignore the detail of instances.
Thus, if the rough structure of cost landscapes like in Figs.~\ref{fig:landscape_fixedbeta}--\ref{fig:landscape_fixedgamma} has some stability against the strength of noises, we can expect a transferability from ideal simulators to noisy real devices.
This point should depend on the detail of devices, so extensive trials and errors on many types of devices will provide some insight towards practical uses.

Most results we report in this paper are provided with fixed $p=8$ QAOA,
but in actual use cases the number of layers should be determined in response to the accuracy, complexity, and noises (on real devices).
The smaller numbers of layers would be interesting towards practical uses.
Indeed, for $p \leq 2$, the linearized QAOA is equivalent to the original QAOA;
thus studying how it transits between $p=2$ and $3$ would be the most important.

In practical point of view, solving penalized constrained problems with parameter transferring QAOA and checking how it is tolerable would be an interesting direction.
Thanks to the transferring, one can omit (or reduce) the optimization for QAOA parameters, so that the total cost gets significantly milder.
Thus we believe that transferring linearized parameters (with combined to a shift technique) contributes to actual use cases of QAOA.

\section*{Acknowledgment}

R.S. thanks Dr. Kazuya Kaneko for a helpful discussion and suggestions in QS11 workshop.

\printbibliography[title=References]

\end{document}